\documentclass[aps,prd,superscriptaddress,showkeys,nofootinbib,twocolumn]{revtex4-2}
\usepackage[english]{babel}
\usepackage{amsmath,bm}
\usepackage{amssymb}
\usepackage{enumitem}
\usepackage{textcomp}
\usepackage{graphicx}
\usepackage{dsfont}
\usepackage{color}

\usepackage[breaklinks=true]{hyperref}
\hypersetup{
	colorlinks=true,
	linkcolor=blue,
	filecolor=magenta,      
	urlcolor=blue,
	citecolor=red,
}

\usepackage{mathrsfs}
\usepackage[normalem]{ulem}

\usepackage{placeins}

\definecolor{dgreen}{rgb}{0,0.6,0}

\begin{document}

	\title{Magnetogenesis in Higgs-Starobinsky inflation}
	
	\author{R.~Durrer}
    \affiliation{D\'{e}partement de Physique Th\'{e}orique, Center for Astroparticle Physics, Universit\'{e} de Gen\`{e}ve, 1211 Gen\`{e}ve 4,  Switzerland}

	\author{O.~Sobol}
	\email{oleksandr.sobol@knu.ua}
	\affiliation{Institute of Physics, Laboratory of Particle Physics and Cosmology, \'{E}cole Polytechnique F\'{e}d\'{e}rale de Lausanne, CH-1015 Lausanne, Switzerland}
    \affiliation{Physics Faculty, Taras Shevchenko National University of Kyiv, 64/13, Volodymyrska Street, 01601 Kyiv, Ukraine}
	
	\author{S.~Vilchinskii}
	\affiliation{D\'{e}partement de Physique Th\'{e}orique, Center for Astroparticle Physics, Universit\'{e} de Gen\`{e}ve, 1211 Gen\`{e}ve 4,  Switzerland}
	\affiliation{Physics Faculty, Taras Shevchenko National University of Kyiv, 64/13, Volodymyrska Street, 01601 Kyiv, Ukraine}

	\date{\today}

	\begin{abstract}
    In the framework of mixed Higgs-Starobinsky inflation, we consider the generation of Abelian gauge fields due to their nonminimal coupling to gravity (in two different formulations of gravity---metric and Palatini). We  couple the gauge-field invariants $F_{\mu\nu}F^{\mu\nu}$ and $F_{\mu\nu}\tilde{F}^{\mu\nu}$ to an integer power of the scalar curvature $R^n$ in Jordan frame and, treating these interactions perturbatively, switch to the Einstein frame where they lead to effective kinetic and axial couplings between gauge fields and inflaton. We determine the power spectra, energy densities, correlation length, and helicality of the generated gauge fields for different values of the nonminimal coupling constants and parameter $n$. We analytically estimate the spectral index $n_{B}$ of the magnetic power spectrum and show that for $n>1$ it is possible to get the scale-invariant or even red-tilted spectrum for a wide range of modes that implies larger correlation length of the generated fields. On the other hand, the magnitude of these fields typically decreases in time becoming very small in the end of inflation. Thus, it is difficult to obtain both large magnitude and correlation length of the gauge field in the frame of this model.
	\end{abstract}

	\maketitle
	
	\section{Introduction}
	\label{sec-In}
Astrophysical observations have established the ubiquity of magnetic fields (MF) of various magnitudes and on various spatial scales in the Universe \cite{Kronberg:1994,Grasso:2001,Widrow:2002,Giovannini:2004,Kandus:2011,Durrer:2013,Subramanian:2016}.
One of the most challenging  problems of modern cosmology is  
the understanding of the origin and evolution of cosmic MF, especially of MF with very large coherence scales of $\lambda_{B}\gtrsim 1\,$Mpc in voids inferred recently through the gamma-ray observations of distant blazars~\cite{Tavecchio:2010,Ando:2010,Neronov:2010,Dolag:2010,Dermer:2011,Taylor:2011,Caprini:2015}. Together with the observations of the cosmic microwave background (CMB) \cite{Planck:2015-pmf,Sutton:2017,Giovannini:2018b,Paoletti:2018} and ultrahigh-energy cosmic rays~\cite{Bray:2018,Neronov:2021} this implies the following constraints on the strength of these fields $10^{-16}\lesssim B_{0}\lesssim 10^{-10}\,$G. 
If the correlation length of the magnetic field is $\lambda_{B}< 1\,$Mpc, the minimal magnetic field strength needed in voids is larger by a factor of $(\lambda_{B}/1\,{\rm Mpc})^{-1/2}$ \cite{Neronov:2010,Caprini:2015}.

The observed intergalactic MF can be explained in the frame of two different scenarios: an astrophysical scenario, where the seed is typically weak and MF is transferred from local sources to larger scales, and, a cosmological (primordial) scenario, where the seed fields are generated prior to galaxy formation in the early Universe and evolves via turbulent decay to achieve correlations scales that can be large by today, see, e.g.,~\cite{Durrer:2013}.

There exist a few purely astrophysical mechanisms, e.g., based on a Biermann battery \cite{Gnedin:2000}, which may generate the seed MF. Further, it can be amplified by adiabatic contraction and different types of dynamos inside astrophysical objects \cite{Zeldovich:1980book,Lesch:1995,Kulsrud:1997,Colgate:2001}. However, it is problematic to transport these  MF out to fill the cosmic voids and to achieve large correlation length. For this reason, a cosmological origin of the large scale MF, as we discuss it in the present paper, seems to be more realistic. 

Many different mechanisms of cosmological seed MF generation have been discussed (for an overview see, for instance, Ref.~\cite{Durrer:2013}). Two basic possibilities for primordial magnetogenesis are inflation and cosmological phase transitions. The latter mechanism typically results in a small coherence length of MF which is of the same order as the horizon size during the transition \cite{Hogan:1983,Quashnock:1989,Vachaspati:1991,Cheng:1994,Sigl:1997,Ahonen:1998}. Moreover, magnetogenesis occurs during the first-order phase transitions while both electroweak and QCD phase transitions in the early Universe are believed to be smooth crossovers. Therefore, inflationary generation of MF seems to be a more attractive scenario. Most probably, during this stage the primordial scalar and tensor perturbations have been generated (see Refs.~\cite{Mukhanov:1992,Durrer:book} for a review). In the similar way, magnetogenesis could occur during inflation by the amplification of quantum vacuum fluctuations of the gauge field and it can lead to arbitrarily large correlation lengths of the  MF observed today.

Since the Maxwell action is conformally invariant, no field amplification occurs in conformally flat expanding Universe unless interaction with the inflaton and/or gravity is introduced. Such possibilities were first discussed in Refs.~\cite{Turner:1988,Ratra:1992,Garretson:1992,Dolgov:1993} and revisited many times in the literature \cite{Giovannini:2001,Bamba:2004,Martin:2008,Kanno:2009,Demozzi:2009,Ferreira:2013,Ferreira:2014,Vilchinskii:2017,Sharma:2017b,Savchenko:2018,Sobol:2018,Shtanov:2020,Talebian:2020,Durrer:2011,Anber:2006,Anber:2010,Barnaby:2012,Caprini:2014,Anber:2015,Ng:2015,Fujita:2015,Adshead:2015,Adshead:2016,Notari:2016,Domcke:2018eki,Cuissa:2018,Shtanov:2019,Shtanov:2019b,Sobol:2019,Kamarpour:2019,Sobol:2021,Domcke:2020zez,Bamba:2008,Bamba:2020,Maity:2021}.
It should be noted that often the form of couplings between the gauge field and gravity or the inflaton field is constructed phenomenologically aiming to obtain the desired magnitude and correlation length of the generated field. The disadvantage of this approach is the absence of physical reasons behind these coupling functions. In some sense, this problem is similar to the choice of the inflaton potential. At present time, there is a great variety of inflationary models (a list of the most popular ones can be found, e.g., in Ref.~\cite{Martin:2013}). In the last decade, many of them have been ruled out because they contradict the CMB anisotropy observations by Planck Collaboration \cite{Planck:2018-infl}. Nevertheless, the number of viable models is still very large. Typically, they belong to the class of plateau models whose potential is concave and becomes very flat for large values of the inflaton. Among them are, e.g., the Starobinsky $R^{2}$ model \cite{Starobinsky:1980}, $\alpha$ attractors \cite{Ferrara:2013,Kallosh:2013}, and Higgs inflation \cite{Bezrukov:2007,Bauer:2008}.

From a physical point of view, inflation models that do not require the introduction of additional fields are the most attractive. One of the examples is the Starobinsky model \cite{Starobinsky:1980} where the usual Einstein-Hilbert action of gravity is extended by a quadratic term $\propto R^{2}$ naturally arising due to quantum corrections. This action can be brought to the standard Einstein gravity with an additional scalar field playing the role of the inflaton. A second example is the Higgs inflationary model \cite{Bezrukov:2007,Bauer:2008}, where inflation occurs due to the Standard Model Higgs boson $h$ nonminimally coupled to spacetime curvature via a term $\propto h^{2}R$  \cite{Bezrukov:2007,Bauer:2008}. In the Einstein frame, the Higgs potential becomes flat at large values of the field and is in accordance with the CMB observations \cite{Planck:2018-infl} for a certain range of the nonminimal coupling constant.
  
The Starobinsky inflation model, based on $R^2$ gravity, and the Higgs inflationary model with a large nonminimal coupling of the Higgs
boson $h$ to the Ricci scalar $R$ lead to the same predictions for the
spectral index of primordial scalar perturbations and to a tensor-to-scalar ratio compatible with the latest Planck data \cite{Planck:2018-infl}. It is natural to consider extensions of these two single-field models to a multifield inflation by combining them.
It was shown \cite{Salvio:2015,Calmet:2016,Salvio:2016,Wang:2017,Ema:2017,St2,Gorbunov:2018,Gundhi:2018,Canko:2019} that such extensions improve the Higgs model that suffer from a strong coupling problem, arising significantly below the Planck scale. The Higgs-Starobinsky inflation model is a theoretically self-consistent model of inflation and reheating below the Planck scale, its predictions are in good  agreements with the Higgs and Starobinsky inflation models as well with  the latest cosmological observations.
  
In the context of the above, it looks natural to use the similar idea of nonminimal coupling in order to realize magnetogenesis in the frame of Starobinsky, Higgs or mixed Higgs-Starobinsky inflation models. The gauge-field generation during the Starobinsky inflation was studied in Ref.~\cite{Savchenko:2018}. A similar analysis for the Higgs inflation model was performed in Refs.~\cite{Kamarpour:2019,Sobol:2021}. In the present work, we consider the mixed Higgs-Starobinsky model of inflation and introduce a more general form of the nonminimal coupling between the gauge field and gravity via the terms $\propto R^n F_{\mu\nu}F^{\mu\nu}$ and $\propto R^n F_{\mu\nu}\tilde{F}^{\mu\nu}$ (here the integer number $n$ is a free parameter of the model). To the best of our knowledge, the gauge-field production in the Higgs-Starobinsky model of inflation have not been studied before.

In the presence of nonminimal coupling of matter fields to the curvature, different formulations of gravity (the usual metric formulation \cite{Einstein:1925} and the Palatini formulation \cite{Palatini:1919}) lead to different physical results. The metric and Palatini formulations of Higgs inflation have been extensively studied in the literature (for a review, see Refs.~\cite{Rubio:2018,Tenkanen:2020,Shaposhnikov:2020}). However, mixed Higgs-Starobinsky inflation has been investigated mostly in the metric formulation in \cite{Salvio:2015,Calmet:2016,Salvio:2016,Wang:2017,Ema:2017,St2,Gorbunov:2018,Gundhi:2018,Canko:2019}, and only a little attention was paid to the Palatini formulation in this model \cite{Enckell:2018,Antoniadis:2018,Gialamas:2019}. In this paper, for completeness, we This is done in Sec.~\ref{sec-HS} where we rederive basic results previously known in the literature and also present some new results concerning the Palatini formulation.
In Sec.~\ref{MG} we consider the gauge field nonminimally coupled to gravity in the Jordan frame and deduce the coupling functions of the gauge field to the inflaton in the Einstein frame, both in the metric and Palatini formulations. In this section some basic equations describing inflationary magnetogenesis are reviewed. In Sec.~\ref{sec-numerical} we analytically investigate the spectral index $n_B$ of generated magnetic fields, which determines the distribution of magnetic energy density among modes with different momenta. Furthermore, numerical results for the power spectrum, energy density, helicity and  correlation length  of the gauge fields generated during the Higgs-Starobinsky  inflation are presented. Section~\ref{sec-concl} is devoted to conclusions. In the Appendix we explain in detail how we perform the Legendre transformation required to bring the action into the canonical Einstein-Hilbert form.

Throughout the work we use the natural units and set $\hbar=c=1$.

	\section{The Higgs-Starobinsky model of inflation}
	\label{sec-HS}
	
	As is well known, in the Jordan frame the action of the Higgs-Starobinsky inflationary model reads as \cite{Ema:2017}:
	\begin{eqnarray}
	\label{a1}
		S[g_{\mu\nu},h]&=&\int d^4x \sqrt{-g}\Big[-\frac{M_p^2}{2}\Big(1+\frac{\xi_h h^2}{M_p^2}\Big)R\nonumber\\
		&&\hspace*{0.5cm}+\frac{\xi_s}{4}R^2+\frac{1}{2}g^{\mu\nu}\partial_\mu h\partial_\nu h-\frac{\lambda}{4}h^4\Big],
	\end{eqnarray}
where $g_{\mu\nu}$ is the spacetime metric with signature $(+,\,-,\,-,\,-)$, $g={\rm det\,}(g_{\mu\nu})$, $R$ is the Ricci curvature scalar, $M_{p}=(8\pi G)^{-1/2}\approx 2.43\times 10^{18}\,{\rm GeV}$ is the reduced Planck mass, $\lambda$ is the dimensionless self-coupling constant of the Higgs (the vacuum expectation value $v$ of the Higgs field is neglected because it is much smaller than the characteristic value of the field $h$ during inflation).
This action  includes the nonminimal coupling of the Higgs field $h$ to the spacetime curvature $R$ which is described by the nonminimal coupling constant $\xi_{h}$, like in Higgs inflationary model \cite{Bezrukov:2007}. 
The action contains also a term quadratic in $R$, described by $\xi_s$ as in the Starobinsky model \cite{Starobinsky:1980}.
In order to bring this action into the canonical Einstein-Hilbert form, one needs to  get rid of the quadratic term by performing a Legendre transform, introducing an additional scalar degree of freedom. Then, one has to perform a conformal transformation of the metric in order to go from the Jordan to the Einstein frame.	
After the Legendre transform (see Appendix) the action has the form \cite{Gorbunov:2018}:
	\begin{eqnarray}
	\label{action-Jordan-prep}
	S[g_{\mu\nu},h,\Psi]&=&\int d^4x \sqrt{-g}\Big[-\frac{M_p^2}{2}\Psi R+\frac{1}{2}\partial_\mu h\partial^\mu h\nonumber\\
	&-&\frac{\lambda}{4}h^4
	-\frac{M_{p}^{4}}{4\xi_{s}}\Big(1-\Psi+\frac{\xi_h h^2}{M_p^2}\Big)^{2}\Big],
	\end{eqnarray}
which is linear in $R$ and contains additional nondynamical scalar degree of freedom $\Psi$
\begin{equation}
		\Psi=1+\frac{\xi_h h^2}{M_p^2}-\frac{\xi_s R}{M_{p}^{2}}.
	\end{equation}	

	With the aim  to get rid of the scalar field $\Psi$ in front of $R$,  a conformal transformation is performed,  which brings the action into the Einstein frame,
	\begin{equation}
		\label{conf-transf}
		g_{\mu\nu}= \frac{1}{\Psi}\bar{g}_{\mu\nu}.
	\end{equation}

	Quite different results follow from the same action in the Jordan frame if one chooses the usual (metric) formulation of gravity or the Palatini one. 
	Depending on whether the metric or the Palatini formulation of gravity is used, the action in the Einstein frame has a different form. 
	
	Let us consider the two cases separately. In both cases we assume that the spacetime metric in the Einstein frame has the Friedmann-Lema\^{i}tre-Robertson-Walker form:
	\begin{equation}
		\bar{g}_{\mu\nu}=\textrm{diag\,}(1,\,-a^{2},\,-a^{2},\,-a^{2}),
	\end{equation}
	where $a=a(t)$ is the scale factor.
Also in both cases the initial conditions and parameters of the inflation effective potentials are constrained using the results of  CMB observations.

\subsection{Einstein frame in the metric formulation}
	
    The main relations and results of this subsection were obtained in \cite{Ema:2017,St2,Gorbunov:2018}, and here we present the results that will be necessary for our investigation  of the  generation of MF in the frame  of Higgs-Starobinsky  model in metric formulation.  
    
	As is well known, in the metric formulation of gravity, the only dynamical degrees of freedom describing the gravitational sector are the components of the metric tensor. All other quantities, such as connection $\Gamma^{\lambda}_{\mu\nu}$, the Riemann tensor $R^{\mu}_{\nu\alpha\beta}$, the Ricci tensor $R_{\mu\nu}$, and the curvature scalar $R$ are expressed in terms of the metric and its derivatives. 
	
	In this case, under a Weyl (or conformal) transformation (\ref{conf-transf}), the Ricci curvature scalar transforms as
	\begin{equation}
	\label{R}
		R=\Psi \Big[\bar{R}-\frac{3}{2\Psi^{2}}\bar{g}^{\mu\nu}\partial_{\mu}\Psi\partial_{\nu}\Psi+3\bar{\nabla}_{\mu}\bar{\nabla}^{\mu}\ln\Psi\Big]
	\end{equation}
	and in the Einstein frame the action takes the form:
	\begin{eqnarray}
		S[\bar{g}_{\mu\nu},h,\Psi] &=&\int d^4x \sqrt{-\bar{g}}\Big[-\frac{M_p^2}{2}\bar{R}\nonumber \\
		&+&\frac{3M_{p}^{2}}{4\Psi^{2}}\bar{g}^{\mu\nu}\partial_{\mu}\Psi\partial_{\nu}\Psi
		+\frac{1}{2\Psi}\bar{g}^{\mu\nu}\partial_\mu h\partial_\nu h\nonumber \\
		&-&\frac{\lambda}{4\Psi^{2}}h^4
		-\frac{M_{p}^{4}}{4\xi_{s}\Psi^{2}}\Big(1-\Psi+\frac{\xi_h h^2}{M_p^2}\Big)^{2}
		\Big],
	\end{eqnarray}
	where  the field $\Psi$ has become dynamical. As a result, we have a two-field inflationary model with the Higgs field $h$ and the scalaron field $\Psi$. Since the latter is not canonically normalized, a new field variable $\phi$ is introduced via
	\begin{equation}
		\Psi=\exp\Big(\sqrt{\frac{2}{3}}\frac{\phi}{M_{p}}\Big)\,.
	\end{equation}
	With this we arrive at an  action with 
	a kinetic mixing of the scalar fields $h$ and $\phi$, which cannot be removed by any field redefinition,
	\begin{eqnarray}
		S[\bar{g}_{\mu\nu},h,\phi]&=&\int d^4 x \sqrt{-\bar{g}}\Big[-\frac{M_p^2}{2}\bar{R}+\frac{1}{2}\bar{g}^{\mu\nu}\partial_{\mu}\phi\partial_{\nu}\phi\nonumber\\
		&+&\frac{1}{2}e^{-\sqrt{\frac{2}{3}}\frac{\phi}{M_{p}}}\bar{g}^{\mu\nu}\partial_\mu h\partial_\nu h-U(\phi,h)\Big].
	\end{eqnarray}
	where the potential $U(\phi,h)$ has the form
	\begin{eqnarray}
		\label{potential-general}
		U(\phi,h)&=&
		\frac{M_{p}^{4}}{4\xi_{s}}\Big[1-e^{-\sqrt{\frac{2}{3}}\frac{\phi}{M_{p}}}\Big(1+\frac{\xi_{h}h^{2}}{M_{p}^{2}}\Big)\Big]^{2}\nonumber\\
		&+&\frac{\lambda h^{4}}{4}e^{-2\sqrt{\frac{2}{3}}\frac{\phi}{M_{p}}}.
	\end{eqnarray}
	
	If one  assumes that during inflation both fields $h$ and $\phi$ are spatially homogeneous and depend only on time, the equations of motion are
	\begin{equation}
	\label{phi1}
		\ddot{\phi}+3H\dot{\phi}=-\partial_\phi U(\phi,h)-\frac{\dot{h}^2}{\sqrt{6}M_{p}}e^{-\sqrt{\frac{2}{3}}\frac{\phi}{M_{p}}},
	\end{equation}
	\begin{equation}
	\label{h1}
		\ddot{h}+3H\dot{h}=-e^{\sqrt{\frac{2}{3}}\frac{\phi}{M_{p}}}\partial_h U(\phi,h)+\sqrt{\frac{2}{3}}\frac{\dot{\phi}}{M_p}\dot{h},
	\end{equation}
	where dot denotes the derivative with respect to physical time $t$ and $H=\frac{\dot{a}}{a}$ is the Hubble parameter. According to the Friedmann equations, it is determined by the total energy density $\rho$ of the Universe
	\begin{equation}
	\label{F}
		H^2=\frac{1}{3M_{p}^2}\rho,
	\end{equation}
	while its derivative depends on the energy density and on the pressure $P$
	\begin{equation}
		\dot{H}=-\frac{1}{2M_{p}^{2}}(\rho+P).
	\end{equation}
	The latter two can be found from the energy-momentum tensor of the scalar fields:
	\begin{eqnarray}
		\hspace*{-0.4cm}&&T_{\mu\nu}=\frac{2}{\sqrt{-g}}\frac{\delta S_{\phi,h}}{\delta \bar{g}^{\mu\nu}}=\partial_{\mu}\phi\partial_{\nu}\phi+e^{-\sqrt{\frac{2}{3}}\frac{\phi}{M_{p}}}\partial_{\mu}h \partial_{\nu}h\nonumber\\ \hspace*{-0.4cm}&&-\bar{g}_{\mu\nu}\Big[\frac{1}{2}\partial_{\alpha}\phi\partial^{\alpha}\phi+\frac{1}{2}e^{-\sqrt{\frac{2}{3}}\frac{\phi}{M_{p}}}\partial_\alpha h\partial^\alpha h-U(\phi,h)\Big],
	\end{eqnarray}
	\begin{equation}
		T_{00}=\rho=\frac{\dot{\phi}^2}{2}+e^{-\sqrt{\frac{2}{3}}\frac{\phi}{M_{p}}}\frac{\dot{h}^2}{2}+U(\phi,h),
	\end{equation}
	\begin{equation}
		T_{ij}=-g_{ij}P, \quad P=\frac{\dot{\phi}^2}{2}+e^{-\sqrt{\frac{2}{3}}\frac{\phi}{M_{p}}}\frac{\dot{h}^2}{2}-U(\phi,h).
	\end{equation}
	
	For further convenience we  compute  the Ricci scalar $R$  in the Jordan frame in terms of the fields $\phi$ and $h$, using (\ref{R})  and  inserting $\bar{R}=-6\dot{H}-12H^{2}=(3P-\rho)/M_{p}^{2}$: 
 \begin{eqnarray}
		R&=&e^{\sqrt{\frac{2}{3}}\frac{\phi}{M_{p}}}\frac{1}{M_{p}^{2}}\Big[3P-\rho-\partial_{\mu}\phi \partial^{\mu}\phi\nonumber\\
		&+&\sqrt{6}M_{p}\frac{1}{\sqrt{-\bar{g}}}\partial_{\mu}\Big(\sqrt{-\bar{g}}\bar{g}^{\mu\nu}\partial_{\nu}\phi\Big)\Big]\nonumber\\
		&=&-\frac{M_{p}^{2}}{\xi_{s}}\Big[e^{\sqrt{\frac{2}{3}}\frac{\phi}{M_{p}}}-1-\frac{\xi_{h}h^{2}}{M_{p}^{2}}\Big].
	\end{eqnarray}

\begin{figure}[ht!]
		\centering
		\includegraphics[width=0.95\linewidth]{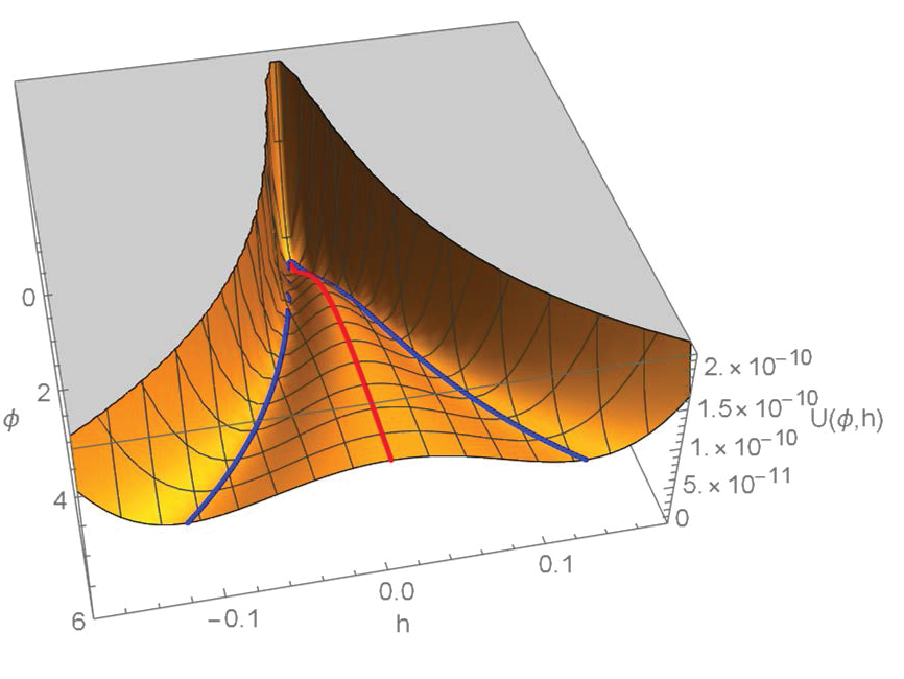}
		\caption{Inflaton potential $U(\phi,h)$ given by Eq.~(\ref{potential-general}). The blue lines show the positions of two valleys while the red line describes the hill at $h=0$.}
		\label{fig-potential}
	\end{figure}

	For the investigation of the inflationary dynamics, 
	we assume for definiteness that the coupling constants are positive, $\xi_{h}$, $\xi_{s}$, $\lambda>0$. The inflation potential $U(\phi,h)$ given by Eq.~(\ref{potential-general}) then has the typical form shown in Fig.~\ref{fig-potential}.
	$U(\phi,h)$ is symmetrical with respect to $h\rightarrow-h$ and has two features: a hill and two valleys. Valleys are formed because the potential has two minima for fixed $\phi>0$ and one for $\phi<0$. For $\phi>0$ the minima are 
	\begin{equation}
		\label{hmin}
		h_{min}^2(\phi)=\frac{M_{p}^2\xi_h}{\lambda\xi_s+\xi_h^2}\Big(e^{\sqrt{\frac{2}{3}}\frac{\phi}{M_{p}}}-1\Big),
	\end{equation}
	while the line $h=0$ corresponds to a maximum (at fixed $\phi>0$).

	Along the valley, the value of the potential is given by the following function of $\phi$ :
	\begin{equation}
		\label{ValleyPot}
		U(\phi,h_{min}(\phi))=\frac{\lambda M_{p}^4}{4 (\lambda\xi_s+\xi_h^2)}\Big(1-e^{-\sqrt{\frac{2}{3}}\frac{\phi}{M_{p}}}\Big)^2.
	\end{equation}
	This effective single-field potential has exactly the same form as the potential in $R^{2}$ Starobinsky inflation or in Higgs inflation.
	
	For the three parameters of the model, $\xi_{h}$, $\xi_{s}$, $\lambda$ one can impose the following constraints \cite{Ema:2017,Gorbunov:2018}:
	\begin{itemize}
		\item in order to remain in the perturbative regime during preheating, 
		\begin{equation}
		\lambda_{\rm eff}=\lambda+\frac{\xi_{h}^{2}}{\xi_{s}}\lesssim 4\pi;
		\end{equation}
		
		\item in order to provide the correct amplitude of primordial scalar perturbations
		\begin{equation}
		\label{constraint}
		 \frac{\xi_{h}^{2}}{\lambda}+\xi_{s}=\frac{N_{\ast}^{2}}{72\pi^{2}A_{s}}\simeq 2\times 10^{9},
		\end{equation}
		where $A_{s}\approx 2\times 10^{-9}$ is the amplitude of the scalar power spectrum, measured by the Planck Collaboration \cite{Planck:2018-infl}, $N_{\ast}\simeq 50-60$ is the number of $e$-foldings before the end of inflation, when the pivot scale of the CMB spectrum crosses the horizon.
	\end{itemize}

	For $\lambda\sim 0.01$ and $\xi_{h}$, $\xi_{s}$ which satisfy the above-mentioned constraints, the inflationary dynamics is effectively the same as in single-field inflation. Indeed, in Ref.~\cite{Ema:2017} it was discussed and confirmed numerically that for $\xi_h \gg 1$ the effective mass of modes orthogonal to the valley $h=\pm h_{min}(\phi)$ is much larger than the Hubble parameter during inflation. As a result, for different initial conditions the Higgs field $h$ quickly rolls down into one of the valleys $\pm h_{min}(\phi)$ and then the slow-roll phase occurs along this valley. As one can easily conclude from Eq.~(\ref{constraint}), unless the parameter $\xi_s$ is fine tuned to take the value $2\times 10^9$ (which corresponds to the case of pure Starobinsky inflation), the parameter $\xi_h$ is much larger than unity. Therefore, for the interesting region in the parameter space the motion orthogonal to the valley is strongly suppressed.
	
	Assuming that relation (\ref{hmin}) holds, the effective action for the field $\phi$ takes the form:
	\begin{eqnarray}
		S_{\phi}&=&\int d^{4}x\sqrt{-\bar{g}}\nonumber\\
		&\times&\Big\{\frac{1}{2}\partial_{\mu}\phi\partial^{\mu}\phi\Big[1+\frac{\xi_{h}}{6(\lambda\xi_s+\xi_{h}^{2})}\Big(1-e^{-\sqrt{\frac{2}{3}}\frac{\phi}{M_{p}}}\Big)^{-1}\Big]\nonumber\\
		&&\hspace*{0.2cm}-\frac{\lambda M_{p}^4}{4 (\lambda\xi_s+\xi_h^2)}\Big(1-e^{-\sqrt{\frac{2}{3}}\frac{\phi}{M_{p}}}\Big)^2\Big\}.
	\end{eqnarray}
	The second term in the square brackets is always subdominant in comparison to unity during inflation. Indeed, the coefficient of this term can be represented as 
	\begin{equation*}
		\frac{\xi_{h}}{6(\lambda\xi_s+\xi_{h}^{2})}=\frac{1}{6\sqrt{\lambda}}\frac{1}{\sqrt{\xi_s+\frac{\xi_{h}^{2}}{\lambda}}}\underset{\leq 1}{\underbrace{\frac{\xi_{h}}{\sqrt{\lambda\xi_s+\xi_{h}^{2}}}}}\leq \frac{3.7\!\times\! 10^{-6}}{\sqrt{\lambda}},
	\end{equation*}
	where the constraint (\ref{constraint}) was used. Thus, for $\lambda\sim 10^{-2}-10^{-3}$ it is much smaller than unity. Therefore, the field $\phi$ is approximately canonically normalized during inflation and
	\begin{eqnarray}
	\label{pot-eff-single}
		S_{\phi}&=&\int d^{4}x\sqrt{-\bar{g}}\Big\{\frac{1}{2}\partial_{\mu}\phi\partial^{\mu}\phi-V(\phi)\Big\},\nonumber\\
		V(\phi)&=&\frac{M_{p}^4}{4 (\xi_s+\xi_h^2/\lambda)}\Big(1-e^{-\sqrt{\frac{2}{3}}\frac{\phi}{M_{p}}}\Big)^2.
	\end{eqnarray}
	For this potential, inflation ends when the slow-roll parameter $\epsilon$ becomes of order unity:
	\begin{eqnarray}
		\epsilon&=&\frac{M_{p}^{2}}{2}\left(\frac{V'}{V}\right)^{2}=\frac{4/3}{\left[\exp\left(\sqrt{\frac{2}{3}}\frac{\phi}{M_{p}}\right)-1\right]^{2}}=1,\nonumber\\
		\phi_{e}&=&M_{p}\sqrt{\frac{3}{2}}\ln\left(1+\frac{2}{\sqrt{3}}\right)\approx 0.94 M_{p}.
	\end{eqnarray}
	During the entire inflation stage, the field is (trans-)Planckian.

	Finally, we  express the curvature scalar in the Jordan frame in terms of $\phi$ during slow roll,
	\begin{equation}
		\label{R-M}
		R=-\frac{M_{p}^2}{ (\xi_s+\xi_h^2/\lambda)}\Big(e^{\sqrt{\frac{2}{3}}\frac{\phi}{M_{p}}}-1\Big).
	\end{equation}

	One can  therefore conclude that for the physically relevant range of the parameters, $\xi_{h}$, $\xi_{s}$, and $\lambda$, the two-field Higgs-Starobinsky model effectively reduces to a single-field model with the effective potential (\ref{ValleyPot}). 
	Moreover, the only combination of the three parameters which is relevant during inflation, $\xi_{s}+\xi_{h}^{2}/\lambda$, is unambiguously fixed by constraint (\ref{constraint}) which follows from CMB observations. In this sense, the slow-roll dynamics in the mixed Higgs-Starobinsky inflation in the metric formulation of gravity is totally equivalent to that of the purely Starobinsky or purely Higgs inflationary models.
	
	\subsection{Einstein frame: Palatini formulation}
	
	In the Palatini formulation of gravity, the components of the connection $\Gamma^{\lambda}_{\mu\nu}$ are considered as independent dynamical variables in addition to the spacetime metric $g_{\mu\nu}$. As a result, the Riemann and Ricci tensors depend only on connection and do not depend on the metric directly. In this case, the curvature scalar $R=g^{\mu\nu}R_{\mu\nu}$ depends trivially on the metric. Consequently, under a Weyl transformation (\ref{conf-transf}) it transforms as
	\begin{equation}
		R=\Psi \bar{R}.
	\end{equation}
	In Einstein frame, the action (\ref{action-Jordan-prep}) now takes the form:
	\begin{eqnarray}
		\label{action-Einstein-P1}
		S[\bar{g}_{\mu\nu},h,\Psi]&=&\int d^4x \sqrt{-\bar{g}}\Big[-\frac{M_p^2}{2}\bar{R}
		+\frac{1}{2\Psi}\bar{g}^{\mu\nu}\partial_\mu h\partial_\nu h\nonumber\\
		&-&\frac{\lambda}{4\Psi^{2}}h^4
		-\frac{M_{p}^{4}}{4\xi_{s}\Psi^{2}}\Big(1-\Psi+\frac{\xi_h h^2}{M_p^2}\Big)^{2}
		\Big].
	\end{eqnarray}
	In contrast to the metric formulation, the scalaron field $\Psi$ does not acquire a kinetic term, i.e., it is not dynamical. The corresponding equation of motion for this field is simply an algebraic constraint,
	\begin{eqnarray}
		\frac{\delta S}{\delta \Psi}&=&-\frac{1}{2\Psi^{2}}\Big[\bar{g}^{\mu\nu}\partial_{\mu}h\partial_{\nu}h-\frac{\lambda h^{4}}{\Psi}\nonumber\\
		&-&\frac{M_{p}^{4}}{\xi_{s}\Psi}\Big(1+\frac{\xi_{h}h^{2}}{M_{p}^{2}}\Big)\Big(1-\Psi+\frac{\xi_{h}h^{2}}{M_{p}^{2}}\Big)\Big]=0.
	\end{eqnarray}
	This allows us to express $\Psi$ in terms of the Higgs field and its derivatives,
	\begin{equation}
		\Psi=
		\frac{1+\frac{2\xi_{h}}{M_{p}^{2}}h^2+\frac{\lambda \xi_s+\xi_h^2}{M_p^4}h^4}{1+\frac{\xi_{h}}{M_{p}^{2}}h^{2} +\frac{\xi_s}{M_p^4}\partial_{\mu}h\partial^{\mu}h }.
	\end{equation}
	Substituting this result back into the action (\ref{action-Einstein-P1}), we finally obtain an action that depends only on the Higgs field:	
	\begin{eqnarray}
		\label{action-Einstein-P2}
		S[\bar{g}_{\mu\nu},h]\!&=&\!\!\int d^4 x \sqrt{-\bar{g}}\bigg\{\!\!-\!\frac{M_p^2}{2}\bar{R}
		\!+\!\frac{1}{\Delta(h)}\Big[\frac{\xi_{s}}{4M_{p}^{4}}(\partial_{\mu}h\partial^{\mu}h)^{2}\nonumber\\
		&+&\frac{1}{2}\Big(1+\frac{\xi_{h} h^{2}}{M_{p}^{2}}\Big)(\partial_{\mu}h\partial^{\mu}h)-\frac{\lambda h^{4}}{4}\Big]
			\bigg\},
	\end{eqnarray}
	where we introduced the shorthand notation \begin{equation*}\Delta(h)=1+\frac{2\xi_{h}}{M_{p}^{2}}h^2+\frac{\lambda \xi_s+\xi_h^2}{M_p^4}h^4.
    \end{equation*}
    
	We now assume that the field $h$ is spatially homogeneous. The equation of motion which then follows from action (\ref{action-Einstein-P2}) is
	\begin{eqnarray}
		\label{eq-Higgs-P}
		&&\ddot{h}\Big(1+\frac{\xi_{h}h^2}{M_{p}^{2}} +\frac{3\xi_s \dot{h}^{2}}{M_p^4}\Big)+3H\dot{h}\Big(1+\frac{\xi_{h}h^2}{M_{p}^{2}} +\frac{\xi_s \dot{h}^{2}}{M_p^4}\Big)\nonumber\\
		&&+\frac{1}{\Delta(h)}\bigg\{\!\!-\!\dot{h}^{2}\Big(1+\frac{\xi_{h}h^2}{M_{p}^{2}} +\frac{3\xi_s \dot{h}^{2}}{M_p^4}\Big)\Big(\frac{\xi_{h}h}{M_{p}^{2}}+\frac{\lambda \xi_s+\xi_h^2}{M_p^4}h^3\Big)\nonumber\\
		&&+\lambda h^{3}\Big(1+\frac{\xi_{h}h^2}{M_{p}^{2}}-\frac{\xi_s \dot{h}^{2}}{M_p^4}\Big)
		\bigg\}=0.
	\end{eqnarray}
	
	The energy-momentum tensor of the Higgs field reads 
	\begin{eqnarray}
		&&T_{\mu\nu}
		=\frac{1}{\Delta(h)}\Bigg\{\!\partial_{\mu}h\,\partial_{\nu}h\Big[1+\frac{\xi_{h} h^{2}}{M_{p}^{2}}+\frac{\xi_{s}}{M_{p}^{4}}(\partial_{\lambda}h\,\partial^{\lambda}h)\Big]\\
		&&-\bar{g}_{\mu\nu}\Big[\frac{\xi_{s}}{4M_{p}^{4}}(\partial_{\!\lambda}h\partial^{\lambda\!}h)^{2}\!+\!\frac{1}{2}\Big(1+\frac{\xi_{h} h^{2}}{M_{p}^{2}}\Big)(\partial_{\!\lambda}h\partial^{\lambda\!}h)\!-\!\frac{\lambda h^{4}}{4}\Big]\!\Bigg\}.\nonumber
	\end{eqnarray}
	For the spatially homogeneous Higgs field, the energy density is
	\begin{equation}
		\rho=\frac{1}{\Delta(h)}\Big[\frac{3\xi_{s}}{4M_{p}^{4}}\dot{h}^{4}+\frac{1}{2}\Big(1+\frac{\xi_{h} h^{2}}{M_{p}^{2}}\Big)\dot{h}^{2}+\frac{\lambda h^{4}}{4}\Big],
	\end{equation}
	while the pressure has the form
	\begin{equation}
		P=\frac{1}{\Delta(h)}\Big[\frac{\xi_{s}}{4M_{p}^{4}}\dot{h}^{4}+\frac{1}{2}\Big(1+\frac{\xi_{h} h^{2}}{M_{p}^{2}}\Big)\dot{h}^{2}-\frac{\lambda h^{4}}{4}\Big].
	\end{equation}
	Trace of the energy-momentum tensor becomes
	\begin{equation}
		T=\bar{g}^{\mu\nu}T_{\mu\nu}=\frac{1}{\Delta(h)}\Big[-\Big(1+\frac{\xi_{h} h^{2}}{M_{p}^{2}}\Big)\dot{h}^{2}+\lambda h^{4}\Big].
	\end{equation}
	Using  the Einstein equation, we immediately obtain the curvature scalar in the Jordan frame
	\begin{equation}
		R=\Psi\bar{R}=
		-\Psi \frac{T}{M_{p}^{2}}=
		\frac{1}{M_{p}^{2}}\frac{\big(1+\frac{\xi_{h} h^{2}}{M_{p}^{2}}\big)\dot{h}^{2}-\lambda h^{4}}{1+\frac{\xi_{h} h^{2}}{M_{p}^{2}}+\frac{\xi_{s}}{M_{p}^{4}}\dot{h}^{2}}.
	\end{equation}
	
	We now perform the slow-roll analysis of the inflation dynamics. For this purpose, we neglect the first term in brackets in action (\ref{action-Einstein-P2}) because it contains higher powers of derivatives (we will justify this approximation a posteriori). Then, introducing a new canonically normalized field $\chi$, we can rewrite the action for the inflaton field in the standard form
	\begin{equation}
		S_{\chi}=\int d^{4}x\,\sqrt{-\bar{g}}\Big[\frac{1}{2}\partial_{\mu}\chi\,\partial^{\mu}\chi-V(\chi)\Big].
	\end{equation}
	Comparing the second term in the brackets of (\ref{action-Einstein-P2}), we find that the new field $\chi$ is related to the Higgs field $h$ by the  differential equation
	\begin{equation}
		\label{rel}
		\frac{d\chi}{dh}=\sqrt{\frac{1+\frac{\xi_{h} h^{2}}{M_{p}^{2}}}{\Delta(h)}}.
	\end{equation}
	The effective potential $V$ is given by the third term in the bracket, 
	\begin{equation}
		V(\chi)=\frac{\lambda h^{4}}{4\Delta(h)}\bigg|_{h=h(\chi)}.
	\end{equation}
	Here $h=h(\chi)$ is the relation between the old and new field found from a solution to Eq.~(\ref{rel}). This relation cannot be obtained analytically without any approximations. Therefore, for now we will keep it in implicit form.
	
	We also compute the slow-roll parameters for this potential ($V'(\chi)= (dV/dh)(d\chi/dh)^{-1}$).
	\begin{equation}
		\epsilon=\frac{M_{p}^{2}}{2}\Big(\frac{V'(\chi)}{V(\chi)}\Big)^{2}=
		\frac{8M_{p}^{2}}{h^{2}}\frac{1+\frac{\xi_{h} h^{2}}{M_{p}^{2}}}{\Delta(h)}\bigg|_{h=h(\chi)},
	\end{equation}
	\begin{eqnarray}
		\eta&=&M_{p}^{2}\frac{V''(\chi)}{V(\chi)}\nonumber\\ &=&
		\!-\frac{8M_{p}^{2}}{h^{2}}\Bigg[1\!+\!\frac{1}{2(1+\frac{\xi_{h} h^{2}}{M_{p}^{2}})}-\frac{3(1+\frac{\xi_{h} h^{2}}{M_{p}^{2}})}{\Delta(h)}\Bigg]\Bigg|_{h(\chi)}\!\!\!.
	\end{eqnarray}
	For $\xi_{h}h^{2}/M_{p}^{2}\gg 1$ these expressions can be simplified as
	\begin{equation}
		\epsilon\simeq \frac{8\xi_{h}}{(\lambda\xi_{s}+\xi_{h}^{2})}\frac{M_{p}^{4}}{h^{4}}, \qquad \eta\simeq -\frac{8M_{p}^{2}}{h^{2}}.
	\end{equation}
	The first slow-roll condition $\epsilon\ll 1$ is satisfied for $h\lesssim M_{p} \Big(\frac{8\xi_{h}}{(\lambda\xi_{s}+\xi_{h}^{2})}\Big)^{1/4}$, while the second slow-roll condition $|\eta|\ll 1$ is violated even earlier, when the Higgs field is still super-Planckian, $h\simeq 2\sqrt{2}M_{p}$. Exactly the same happens in pure Higgs inflation in the Palatini formulation, in which
	the potential has the form $V_{0}(\chi)=C\,\tanh^{4}(\sqrt{\xi_{h}}\chi)$ and a simple relation $h(\chi)=\frac{M_{p}}{\sqrt{\xi_{h}}}\sinh(\sqrt{\xi_{h}}\chi/M_{p})$ exists. The slow-roll parameters have the form \cite{Sobol:2021}:
	\begin{eqnarray}
		\epsilon_{0}&=&\frac{32\xi_{h}}{\sinh^{2}(2\sqrt{\xi_{h}}\chi/M_{p})}=\frac{8M_{p}^{2}}{h^{2}}\frac{1}{1+\frac{\xi_{h}h^{2}}{M_{p}^{2}}},\\
		\eta_{0}&=&\frac{16\xi_{h}(4-\cosh(2\sqrt{\xi_{h}}\chi/M_{p}))}{\sinh^{2}(2\sqrt{\xi_{h}}\chi/M_{p})}\nonumber\\
		&=&-\frac{8M_{p}^{2}}{h^{2}}\Bigg[1-\frac{5}{2(1+\frac{\xi_{h} h^{2}}{M_{p}^{2}})}\Bigg].
	\end{eqnarray}
	
	Since the inequality $\xi_{h}h^{2}/M_{p}^{2}\gg 1$ holds during the whole inflationary stage (if $\xi_{h}\gg 1$), we can now simplify the relation (\ref{rel}) between $\chi$ and $h$:
	\begin{equation}
		\frac{d\chi}{dh}\simeq\frac{M_{p}}{h}\sqrt{\frac{\xi_{h}}{\lambda \xi_s+\xi_h^2}},
	\end{equation}
	which  leads to
	\begin{equation}
		\chi=M_{p} \sqrt{\frac{\xi_{h}}{\lambda \xi_s+\xi_h^2}} \ln \frac{h}{M_{p}}+\chi_{0},
	\end{equation}
	where $\chi_{0}$ is a constant. Hence,
	\begin{equation}
		h=M_{p}\exp\Big[\sqrt{\frac{\lambda \xi_s+\xi_h^2}{\xi_{h}}}\frac{(\chi-\chi_{0})}{M_{p}}\Big],
	\end{equation}
	and the effective potential takes the form
	\begin{equation}
		\!\!V(\chi)\!\simeq\!\frac{\lambda M_{p}^{4}}{4(\xi_{h}^{2}+\lambda\xi_{s})}\Big[1-8\exp\Big(\!-2\sqrt{\frac{\lambda \xi_s+\xi_h^2}{\xi_{h}}}\frac{\chi}{M_{p}}\Big)\Big].\label{pot-approx}
	\end{equation}
	Here, we have chosen the value of the constant $\chi_{0}$ such that the potential has the same asymptotic behavior as in the case of pure Higgs inflation (in the Palatini formalism):
	\begin{eqnarray}
		V_0(\chi)&=&\frac{\lambda M_{p}^{4}}{4\xi_{h}^{2}}\tanh^{4}\Big(\frac{\sqrt{\xi_{h}}\chi}{M_{p}}\Big)\nonumber\\
		&\simeq& \frac{\lambda M_{p}^{4}}{4\xi_{h}^{2}}\Big[1-8\exp\Big(-2\frac{\sqrt{\xi_{h}}\chi}{M_{p}}\Big)\Big].
	\end{eqnarray}
	
	The effective potential (\ref{pot-approx}) belongs to the class of plateau inflationary models.
	
	When we estimate the amplitude of primordial scalar perturbations generated during inflation with the potential (\ref{pot-approx}), 
we conclude that it depends only on $\xi_{h}$ and $\lambda$ and does not depend on $\xi_{s}$:
	\begin{equation}
		A_{s}=\frac{N_{\ast}^{2}}{12\pi^{2}}\frac{\lambda}{\xi_{h}}, \quad \Rightarrow \quad \frac{\xi_{h}}{\lambda}=\frac{N_{\ast}^{2}}{12\pi^{2}A_{s}}\approx 10^{10}.
		\label{constraint-P}
	\end{equation} 
	This  constraint  is the  analog of the constraint (\ref{constraint}). Thus, for $\lambda=10^{-2}-10^{-3}$ we can fix $\xi_{h}=10^{8}-10^{7}\gg 1$.
	
	For completeness, we also present the spectral index of scalar perturbations and the tensor-to-scalar ratio predicted by the Higgs-Starobinsky model in Palatini formulation:
	\begin{eqnarray}
	    n_{s}&\simeq& 1-\frac{2}{N_{\ast}}\approx 0.96,\nonumber\\
	    r&\simeq&\frac{2}{\xi_{h}N_{\ast}^{2}}\Big(1+\frac{\lambda\xi_{s}}{\xi_{h}^{2}}\Big)^{-1}<10^{-10}.
	\end{eqnarray}

Let us now simplify the equations describing inflation using the slow-roll approximation. As we already saw above, the slow-roll parameter $\eta$ becomes of order unity for $h\simeq 2\sqrt{2}M_{p}$. For super-Planckian fields, and for $\xi_{h}\gg 1$ the Hubble parameter is almost constant and equal to
	\begin{equation}
		\label{Hubble-P}
		H\approx M_{p}\sqrt{\frac{\lambda}{12(\lambda\xi_{s}+\xi_{h}^{2})}}={\rm const}.
	\end{equation}
	
	Equation (\ref{eq-Higgs-P}) takes the form:
	\begin{equation}
		\label{eq-Higgs-P2}
		\ddot{h}+3H\dot{h}+\frac{\lambda}{\lambda\xi_{s}+\xi_{h}^{2}}\frac{M_{p}^{4}}{h}=0.
	\end{equation}
	Neglecting the second derivative (again, due to slow roll), we can estimate the inflaton velocity
	\begin{equation}
		\dot{h}\approx -\frac{\lambda}{\lambda\xi_{s}+\xi_{h}^{2}}\frac{M_{p}^{4}}{3Hh}=-\frac{2}{\sqrt{3}}\sqrt{\frac{\lambda}{\lambda\xi_{s}+\xi_{h}^{2}}}\frac{M_{p}^{3}}{h}.
	\end{equation}
	This estimate allows us to make sure that  neglecting the terms $\sim \dot{h}^{4}$ in the action, is self-consistent during inflation. This conclusion follows from the computation:
	\begin{multline}
		\frac{\frac{\xi_{s}}{4M_{p}^{4}}(\partial_{\mu}h\partial^{\mu}h)^{2}}{\frac{1}{2}\Big(1+\frac{\xi_{h} h^{2}}{M_{p}^{2}}\Big)(\partial_{\mu}h\partial^{\mu}h)}\approx \frac{\xi_{s}\dot{h}^{2}}{2\xi_{h}M_{p}^{2}h^{2}}\\
		\approx \frac{2}{3\xi_{h}}\times \underset{\lesssim 1}{\underbrace{\Big(\frac{M_{p}}{h}\Big)^{4}}}\times \underset{\leq 1}{\underbrace{\frac{\lambda\xi_{s}}{\lambda\xi_{s}+\xi_{h}^{2}}}}\ll 1,
	\end{multline}

	The curvature scalar in the Jordan frame can also be simplified in the slow-roll approximation:
	\begin{equation}
		\label{R-P}
		R\approx -\frac{\lambda}{\xi_{h}}h^{2}.
	\end{equation}
	Note that the prefactor in this expression is fully fixed by the constraint (\ref{constraint-P}).
	
	We also want to study the impact of the parameter $\xi_{s}$ on the inflationary dynamics. Since the Hubble parameter is almost constant, it is convenient to introduce the dimensionless time
	\begin{equation}
		\tilde{t}=Ht=\sqrt{\frac{\lambda}{12(\lambda\xi_{s}+\xi_{h}^{2})}}M_{p}t.
	\end{equation}
	In terms of $\tilde t$, Eq.~(\ref{eq-Higgs-P2}) reads 
	\begin{equation}
		h''(\tilde{t})+3h'(\tilde{t})+\frac{12M_{p}^{2}}{h}=0,
	\end{equation}
	where prime denotes the derivative with respect to $\tilde{t}$. The initial conditions for this equation leading to our solution are 
	\begin{equation}
		h(0)=h_{0}, \quad h'(0)=-4M_{p}^{2}/h_{0}.
	\end{equation}
	
	We conclude that during the Higgs-Starobinsky inflation in the Palatini formalism only two parameters, $\lambda$ and $\xi_{h}$, are constrained from the CMB observations, see Eq.~(\ref{constraint-P}). The third parameter, $\xi_{s}$ is free. Its impact on the inflationary dynamics is rather trivial: it changes the characteristic timescale $t_{H}=H^{-1}$ and the energy scale $\rho_{\rm inf}=3H^{2}M_{p}^{2}$ during inflation, where $H$ is given by (\ref{Hubble-P}).

	\section{Magnetogenesis from nonminimal coupling to gravity}
	\label{MG}
	
	Let us now consider the Abelian gauge field nonminimally coupled to gravity in the Jordan frame. The corresponding part of the action has the form:
	\begin{eqnarray}
		\label{action-GF-gen}
		S_{\mathrm{GF}}&=&\int d^{4}x\sqrt{-g}\Big[-\frac{1}{4}F_{\mu\nu}F^{\mu\nu}\nonumber\\
		&-&\frac{(-R)^{n}}{2M_{p}^{2n}}\Big(\kappa_{1}F_{\mu\nu}F^{\mu\nu}+\kappa_{2}F_{\mu\nu}\tilde{F}^{\mu\nu}\Big)\Big],
	\end{eqnarray}
	where the first term is the free gauge (Maxwell) Lagrangian while the second and third terms describe the kinetic and axial nonminimal coupling with the corresponding dimensionless coupling constants $\kappa_{1}$ and $\kappa_{2}$ respectively. Note that in our convention for the metric signature, $R=-6(\dot{H}+2H^2)<0$ is always negative during inflation. Then, the effective gauge coupling of any charged matter field is rescaled by a factor of $[1+2\kappa_1 (-R)^n/M_{p}^{2n}]^{-1}<1$ for positive $\kappa_1$. Therefore, in order to avoid the strong coupling problem, we fix the sign of $\kappa_1$ to be positive. On the other hand, the sign of $\kappa_2$ may be chosen arbitrarily. For the sake of definiteness, we also take it positive. The integer $n$ is a free parameter of  model. The case with $n=1$ for purely Starobinsky inflation was considered in Ref.~\cite{Savchenko:2018}, while for purely Higgs inflation it was studied in Ref.~\cite{Sobol:2021}.
	
	Note that there are also other ways to couple the gauge field to gravity, e.g., by its coupling to $R_{\mu\nu}$ or $R_{\mu\nu\alpha\beta}$ with different possibilities to contract the indices. Moreover, it is even \textit{necessary} to consider the combination of terms with $R$, $R_{\mu\nu}$, and $R_{\mu\nu\alpha\beta}$ with specific coefficients in order to avoid the ghost instabilities in the theory (see, e.g., Ref.~\cite{Jimenez:2013}). However, during inflation, the equation of state is close to vacuumlike, $P\simeq -\rho$, where the difference between these cases is proportional to the slow-roll parameter $\epsilon$. Therefore, $T_{\mu\nu}\approx \rho g_{\mu\nu}$, which immediately implies from the Einstein equations that $R_{\mu\nu}\propto R g_{\mu\nu}$. Moreover, for de Sitter spacetime $R_{\mu\nu\alpha\beta}\propto R(g_{\mu\alpha}g_{\nu\beta}-g_{\mu\beta}g_{\nu\alpha})$. These relations reduce all possible couplings to a single case in any given order in $R$, which is represented by Eq.~(\ref{action-GF-gen}).
	
	We consider only the case when the backreaction of the gauge field on the evolution of the scalar fields and scale factor during inflation can be neglected. This allows us to use the same conformal transformation of the metric as in the previous section in order to switch to the Einstein frame. Moreover, we may use the unperturbed Friedmann and Klein-Gordon equations in order to describe the evolution of the Universe during inflation.
	
	In the Einstein frame, we obtain the following gauge-field action:
	\begin{equation}
		\label{action-GF}
		S_{\mathrm{GF}}=-\frac{1}{4}\int d^{4}x\sqrt{-\bar{g}}\Big[I_{1}F_{\mu\nu}F^{\mu\nu}+I_{2}F_{\mu\nu}\tilde{F}^{\mu\nu}\Big],
	\end{equation}
	where all contractions are now performed by means of the new metric $\bar{g}_{\mu\nu}$. The kinetic and axial coupling functions $I_{1,2}$ have the form:
	\begin{equation}
		I_{j}=\delta_{1j}+\frac{2\kappa_{j}}{M_{p}^{2n}} (-R)^{n}.
	\end{equation}

	Using the results of the previous section, Eqs.~(\ref{R-M}) and (\ref{R-P}), we can express the coupling functions as follows:
	\begin{equation}\label{e:Ijmetric}
		I_{j}=\delta_{1j}+2\kappa_{j}\Big(\frac{72\pi^{2} A_{s}}{N_{\ast}^{2}}\Big)^{n} \Big(e^{\sqrt{\frac{2}{3}}\frac{\phi}{M_{p}}}-1\Big)^{n}
	\end{equation}
	for the metric formulation of gravity and
	\begin{equation}\label{e:Ijpala}
		I_{j}=\delta_{1j}+2\kappa_{j}\Big(\frac{12\pi^{2} A_{s}}{N_{\ast}^{2}}\Big)^{n} \Big(\frac{h}{M_{p}}\Big)^{2n}
	\end{equation}
	for the Palatini formulation, where we work with the Higgs field even though it is not canonically normalized.
	Note that $R$ depends on the parameters $\xi_{h}$, $\xi_{s}$, and $\lambda$ only in the form of the combination which can be fixed from the amplitude of the primordial power spectrum $A_{s}$ and the number of $e$-foldings $N_*$. In Eqs.~(\ref{e:Ijmetric})--(\ref{e:Ijpala}) we already used these constraints and expressed $R$ in terms of $A_{s}$ and $N_*$ which are determined from the CMB observations \cite{Planck:2018-infl}.
	
	From Eqs.~(\ref{e:Ijmetric}) and (\ref{e:Ijpala}) one can conclude that, effectively, not $\kappa_{1}$ and $\kappa_2$ play the role of coupling constants but rather the products $\kappa_{1,2} c_{0}^{n}$, where $c_0=12\pi^2 A_s/N_{\ast}^{2}\simeq 10^{-10}$. Therefore, in order to be in a perturbative regime, one must make sure that $|\kappa_{1,2} c_0^n|\ll 1$. However, this is not a strong constraint on the possible values of the coupling constants $\kappa_1$ and $\kappa_2$ because $c_0$ is very small. In particular, as we will see in Sec.~IV, effective gauge-field production can occur for the values of $\kappa_{1,2}$ which are a few orders of magnitude smaller than the upper bound $c_0^{-n}$.

Let us now derive the set of equations which describe the generation of gauge fields during inflation. First of all, let us emphasize that all our formalism is applicable only in the case when the generated fields do not backreact on the inflaton dynamics. If this is not the case, the gauge field must be treated self-consistently from the very beginning (i.e., we have to take it into account when performing the conformal transformation to the Einstein frame and when determining the expansion dynamics of the Universe). Thus, we restrict ourselves to values of the parameters $\kappa_{1,2}$ and $n$ for which backreaction can be neglected.

In the case with both kinetic and axial coupling functions described by action (\ref{action-GF}), the equations of motion for the electric and magnetic fields (defined as $F^{0i}=\frac{1}{a} E^{i}$, $F^{ij}=\frac{1}{a^{2}}\varepsilon^{ijk}B^{k}$) have the form:
\begin{eqnarray}
	&&\dot{\mathbf{E}}+2H\mathbf{E}-\frac{1}{a}{\rm rot\,}\mathbf{B}+\frac{\dot{I}_{1}}{I_{1}}\mathbf{E}+\frac{\dot{I}_{2}}{I_{1}}\mathbf{B}=0,\label{Maxwell-2}\\
	&&\dot{\mathbf{B}}+2H\mathbf{B}+\frac{1}{a}{\rm rot\,}\mathbf{E}=0,\\
	&&{\rm div\,}\mathbf{E}=0,\qquad {\rm div\,}\mathbf{B}=0.
\end{eqnarray}

In the Coulomb gauge where ${\rm div\,}\mathbf{A}=0$ and $A_{\mu}=(0,\,\mathbf{A})$, electric 
and magnetic components of the gauge field are expressed as follows:
\begin{equation}
	\label{fields-E-and-B}
	\mathbf{E}=-\frac{1}{a}\dot{\mathbf{A}}, \qquad \mathbf{B}=\frac{1}{a^{2}} {\rm rot\,}\mathbf{A}.
\end{equation}
With such a substitution, the last three of the Maxwell equations are identically satisfied and Eq.~(\ref{Maxwell-2}) reads as
\begin{equation}
	\label{Maxwell-A}
	\ddot{\mathbf{A}}+\left(H+\frac{\dot{I}_{1}}{I_{1}}\right)\dot{\mathbf{A}}-\frac{1}{a^{2}}\partial_{i}^{2}\mathbf{A}-\frac{1}{a}\frac{\dot{I}_{2}}{I_{1}}{\rm rot\,}\mathbf{A}=0.
\end{equation}

In order to study the gauge field generation by amplification of quantum fluctuations, we consider the quantum gauge field operator, decomposed over the full set of creation (annihilation) operators $\hat{b}^{\dagger}_{\mathbf{k},\lambda}$ ($\hat{b}_{\mathbf{k},\lambda}$) of the modes with momentum $\mathbf{k}$ and transverse circular polarization $\lambda$
\begin{multline}
	\label{quant-operator}
	\hat{\mathbf{A}}(t,\mathbf{x})=\int\frac{d^{3}\mathbf{k}}{(2\pi)^{3/2}}\!\!\sum_{\lambda=\pm}\Big\{ \boldsymbol{\varepsilon}_{\lambda}(\mathbf{k})\hat{b}_{\mathbf{k},\lambda}A_{\lambda}(t,\mathbf{k})e^{i\mathbf{k}\cdot\mathbf{x}}\\
	+\boldsymbol{\varepsilon}^{*}_{\lambda}(\mathbf{k})\hat{b}^{\dagger}_{\mathbf{k},\lambda}A^{*}_{\lambda}(t,\mathbf{k})e^{-i\mathbf{k}\cdot\mathbf{x}}\Big\},
\end{multline}
where the polarization three-vectors $\boldsymbol{\varepsilon}_{\lambda}(\mathbf{k})$ have the following properties:
\begin{eqnarray}
	\hspace*{-0.3cm}&&\mathbf{k}\cdot\boldsymbol{\varepsilon}_{\lambda}(\mathbf{k})=0,\qquad \boldsymbol{\varepsilon}^{*}_{\lambda}(\mathbf{k})=\boldsymbol{\varepsilon}_{-\lambda}(\mathbf{k}),\nonumber\\
	\hspace*{-0.3cm}&&[i\mathbf{k}\times\boldsymbol{\varepsilon}_{\lambda}(\mathbf{k})]=\lambda k \boldsymbol{\varepsilon}_{\lambda}(\mathbf{k}), \quad \boldsymbol{\varepsilon}^{*}_{\lambda}(\mathbf{k})\cdot\boldsymbol{\varepsilon}_{\lambda'}(\mathbf{k})=\delta_{\lambda\lambda'}. 
\end{eqnarray}
The creation and annihilation operators satisfy the canonical commutation relations
\begin{equation}
	[\hat{b}_{\lambda,\mathbf{k}},\,\hat{b}^{\dagger}_{\lambda',\mathbf{k}'}]=\delta_{\lambda\lambda'}\delta^{(3)}(\mathbf{k}-\mathbf{k}').
\end{equation}

In the absence of backreaction, the Fourier decomposition (\ref{quant-operator}) is convenient because the operator $\hat{\mathbf{A}}$ satisfies a linear equation of motion. Therefore, each gauge-field mode evolves independently. The corresponding equation of motion for the mode function can be derived from Eq.~(\ref{Maxwell-A}):
\begin{multline}
	\label{eq-mode-1}
	\ddot{A}_{\lambda}(t,\mathbf{k})+\left(H+\frac{\dot{I}_{1}}{I_{1}}\right)\dot{A}_{\lambda}(t,\mathbf{k})\\+\left(\frac{k^{2}}{a^{2}}-\lambda\frac{k}{a}\frac{\dot{I}_{2}}{I_{1}}\right)A_{\lambda}(t,\mathbf{k})=0.
\end{multline}

For further convenience, we introduce the new functions
\begin{equation}
\mathcal{B}_{\lambda}(t,\mathbf{k})=\sqrt{2kI_{1}}A_{\lambda}(t,\mathbf{k}), \quad 
\mathcal{E}_{\lambda}=\sqrt{2kI_{1}}(a/k)\dot{A}_{\lambda}.
\end{equation}
Then, in terms of $\mathcal{B}_{\lambda}$, Eq.~(\ref{eq-mode-1}) takes the form 
\begin{multline}
	\label{eq-mode-2}
	\ddot{\mathcal{B}}_{\lambda}(t,\mathbf{k})+H\dot{\mathcal{B}}_{\lambda}(t,\mathbf{k})\\
	+\bigg(\frac{k^{2}}{a^{2}}\!-\!\frac{H\dot{I}_{1}}{2I_{1}}\!-\!\frac{1}{\sqrt{I_{1}}}\frac{d^{2}\!\sqrt{I_{1}}}{dt^{2}}\!-\!\lambda\frac{k}{a}\frac{\dot{I}_{2}}{I_{1}}\bigg)\mathcal{B}_{\lambda}(t,\mathbf{k})=0.
\end{multline}
In conformal time $\tau=\int^{t} dt'/a(t')$, Eq.~(\ref{eq-mode-2}) can be rewritten as
\begin{equation}
	\label{eq-mode-conf}
	\frac{d^2\! \mathcal{B}_\lambda(\tau,\,\mathbf{k})}{d\tau^2}+\!\bigg(k^{2}\!-\!\frac{1}{\sqrt{I_{1}}}\frac{d^{2}\!\sqrt{I_{1}}}{d\tau^{2}}\!-\!\frac{\lambda k}{I_{1}}\frac{dI_2}{d\tau}\bigg)\mathcal{B}_{\lambda}(\tau,\mathbf{k})=0.
\end{equation}
There are two important limiting cases in these equations. First, if $k|\tau|\gg 1$ (the mode is far inside the horizon), the first term in brackets in Eq.~(\ref{eq-mode-conf}) dominates over the last two ones. In this regime, the mode function is described by the Bunch-Davies vacuum solution
\begin{equation}
	\label{Bunch-Davies-vacuum}
	\mathcal{B}_{\lambda}(\tau,\mathbf{k})=e^{-ik\tau}, \quad k\tau\to-\infty.
\end{equation}
Therefore, we observe that inside the horizon, the mode just oscillates in time without a significant change of its amplitude. It does not feel the impact of the kinetic and axial coupling functions and, thus, must not be taken into account in the generated field. In other words, modes inside the horizon correspond to unamplified vacuum fluctuations.

In the opposite case, when $k|\tau|\ll 1$, the mode does not oscillate any more and its evolution is determined by three last terms in brackets in Eq.~(\ref{eq-mode-2}) [or two last terms in Eq.~(\ref{eq-mode-conf})]. Since these terms depend on coupling functions, we conclude that outside the horizon, evolution of the mode function is different from the Bunch-Davies vacuum solution. In particular, the mode amplification may occur only in this regime. Only modes outside the horizon should be taken into account when computing the characteristics of the generated gauge field.

The last mode which must be taken into account is a matter of convention. The natural expression for the threshold momentum can be deduced from the requirement that the first term in brackets in Eq.~(\ref{eq-mode-2}) becomes equal to the contribution of the kinetic coupling (the second and third terms) or the axial coupling (the fourth term). Such a momentum, $k_{h}$, in what follows will be associated with the horizon crossing (this is an effective gauge-field horizon, not to be confused with the Hubble horizon $k_{\rm H}=aH$). It has the following expression:
\begin{equation}
	\label{k-h}
	k_{h}(t)=\max\{k_{1}(t), \, k_{2}(t)\},
\end{equation}
where
\begin{equation}
	\label{k-h1}
	k_{1}(t)=
	\underset{t'\leq t}{\max} \Big\{a(t')\Big|\frac{H(t')}{2I_1(t')}\frac{dI_1}{dt'}\!+\!\frac{1}{\sqrt{I_{1}(t')}}\frac{d^{2}\!\sqrt{I_{1}}}{dt^{'2}}\Big|^{\frac{1}{2}}\Big\},
\end{equation}
\begin{equation}
	\label{k-h2}
	k_{2}(t)=
	\underset{t'\leq t}{\max}\Big\{a(t')\Big|\frac{1}{I_{1}(t')}\frac{dI_{2}(t')}{dt'}\Big|\Big\}.
\end{equation}
Thus, the physically relevant modes at the moment $t$ are those which have crossed the horizon from the beginning of inflation until the moment of time $t$. 

The spectral densities of the electric and magnetic energy densities are expressed in terms of mode functions as follows:
\begin{eqnarray}
	P_{B}(k)\equiv\frac{d\rho_{B}}{d\ln \,k}&=&\sum_{\lambda=\pm} \frac{k^{5}}{4\pi^{2}a^{4}}I_{1}|A_{\lambda}(t,k)|^{2}\nonumber\\&=&\sum_{\lambda=\pm} \frac{k^{4}}{8\pi^{2}a^{4}}|\mathcal{B}_{\lambda}(t,k)|^{2}, \label{en-dens-sp-B}\\
	P_{E}(k)\equiv\frac{d\rho_{E}}{d\ln \,k}&=&\sum_{\lambda=\pm} \frac{k^{3}}{4\pi^{2}a^{2}}I_{1}|\dot{A}_{\lambda}(t,k)|^{2}\nonumber\\
	&=&\sum_{\lambda=\pm} \frac{k^{4}}{8\pi^{2}a^{4}}|\mathcal{E}_{\lambda}(t,k)|^{2}.\label{en-dens-sp-E}
\end{eqnarray}
Note that for the free gauge field both spectral densities coincide $P_{B,0}(k)=P_{E,0}(k)= 2P_{0}(k)$, where
\begin{equation}
    \label{P0}
    P_{0}(k)\equiv\frac{k^{4}}{8\pi^{2}a^{4}}
\end{equation}
is the power spectrum of one given circular polarization.

The energy densities of generated fields can be found by integrating the spectral densities over the range of physically relevant modes, i.e., those with momenta $k\leq k_{h}(t)$:
\begin{eqnarray}
\label{en-dens}
	\rho_{E}(t)&=&\frac{\langle E^2\rangle}{2}=
	\int_{0}^{k_{h}(t)}\frac{dk}{k} P_{E}(k),\nonumber\\
	\rho_{B}(t)&=&\frac{\langle B^2\rangle}{2}=
	\int_{0}^{k_{h}(t)}\frac{dk}{k} P_{B}(k).
\end{eqnarray}
Knowing the magnetic power spectrum  we can determine the correlation length of the generated magnetic field \cite{Durrer:2013}
\begin{eqnarray}
	\label{lambda-B}
	\lambda_{B}&=&\left<\frac{2\pi a}{k}\right>\!\!=\frac{1}{\rho_{B}}\int_{0}^{k_{h}(t)}\frac{dk}{k}\frac{2\pi a}{k}P_{B}(k)\nonumber\\
	&=&\frac{1}{\rho_{B}}\int\limits_{0}^{k_{h}(t)}\!\!\!dk\,\frac{k^{2}}{4\pi a^{3}}\big\{|\mathcal{B}_{+}(t,k)|^{2}+|\mathcal{B}_{-}(t,k)|^{2} \big\}.
\end{eqnarray}
Note that the definition of the correlation length may be not unique; however, our choice is rather natural as it gives the average wavelength of gauge-field modes weighted by the magnetic spectral energy density.

For nonzero $\kappa_{2}$, Eq.~(\ref{eq-mode-2}) depends on $\lambda$, meaning that modes with different circular polarizations evolve differently. As a result, the generated fields are helical, i.e., they have nonzero helicity
\begin{equation}
	\label{helicity}
	\mathcal{H}\!=\!\frac{I_{1}}{a}\langle\mathbf{A}\cdot\mathbf{B}\rangle\!=\!\!\!\!\!\int\limits_{0}^{k_{h}(t)}\!\!\!\frac{dk\ k^{2}}{4\pi^{2} a^{3}}\big\{|\mathcal{B}_{+}(t,k)|^{2}\!-\!|\mathcal{B}_{-}(t,k)|^{2} \big\}.
\end{equation}
Comparing Eqs.~(\ref{lambda-B}) and (\ref{helicity}), we can deduce the so-called realizability condition:
\begin{equation}\label{e:realize}
	|\mathcal{H}|\leq \frac{\lambda_{B}\,\rho_{B}}{\pi}.
\end{equation} 
The equality in \eqref{e:realize} is reached when one circular polarization completely dominates over the second one. Such gauge fields are called maximally helical. 
In general, the degree of ``helicality'' can be characterized by the quantity
\begin{equation}
\label{helicality}
	\eta_{h}=\frac{\pi |\mathcal{H}|}{\lambda_{B} \, \rho_{B}}, \qquad 0\leq \eta_{h}\leq 1.
\end{equation}

\section{Results and discussion}
\label{sec-numerical}

In Sec.~\ref{sec-HS} we noted  that the Higgs-Starobinsky model in the metric formalism during inflation reduces to an effective single-field theory which is equivalent to pure Starobinsky or pure Higgs inflation. In the Palatini formalism it is not the case, however, the difference from pure Higgs inflation is trivial: in order to reduce one model to the other one needs simply to rescale the time and the energy. In this work, we do not consider the case of coupling between gauge fields and curvature with $n=1$ which was already considered in Ref.~\cite{Sobol:2021} for the case of Higgs inflation both in the metric and Palatini formalisms. We  instead consider the cases with $n=2$ and $n>2$. The motivation for this is that we expect to find a spectrum of the generated gauge fields that is less blue-tilted than in the case $n=1$ and, as a result, reach  a larger correlation length of the magnetic fields. In what follows we check this expectation.

In both cases, metric and Palatini, our algorithm is the same. 
We solve numerically the background Friedman equations (\ref{F}) together with  the Klein-Gordon equations (\ref{phi1})-(\ref{h1}) (in the metric case) or 
(\ref{eq-Higgs-P}) (in the Palatini case)  which do not take into account the gauge field. Thus, we neglect backreaction of the generated fields on the background evolution which sets a bound on the applicability of our approach given by   the condition
\begin{equation}
	\label{weak-field}
	I_{1}\langle \mathbf{E}^{2}\rangle,\ I_{1}\langle \mathbf{B}^{2}\rangle\ll \rho_{\rm inf}
\end{equation}
during the entire inflationary stage. It should  be  noted that only in this case we are allowed to use the linear approximation and to consider the generation of the gauge field on the inflation background determined by the unperturbed Friedmann and Klein-Gordon equations.
If this condition were violated, all our perturbative treatment of the gauge field nonminimally coupled to gravity becomes incorrect. In this case, one needs to take into account the presence of the gauge field already at the level of conformal transformation to the Einstein frame. This would lead to a more complicated nonlinear action for the gauge field. This requires a self-consistent study of the joint evolution of the inflaton(s) and gauge fields from the very beginning which is beyond the scope of the present study.

When the time dependences of the scale factor and the inflaton field are determined, 
we choose different values of coupling parameters  $\kappa_{1}$ and $\kappa_{2}$ and solve numerically the mode equation (\ref{eq-mode-2}) with the boundary condition (\ref{Bunch-Davies-vacuum}) for all physically relevant modes (those which are inside the horizon at the beginning of our simulation and cross the horizon during inflation).

After that, we calculate the electric and magnetic spectral densities (\ref{en-dens-sp-B})--(\ref{en-dens-sp-E}). 
Integrating them over the modes which have already crossed the horizon at a given time $t$, we obtain the  energy densities $\rho_{E,B}(t)$, the magnetic correlation length $\lambda_{B}$, 
and the helicality $\eta_{h}$ using Eqs.~(\ref{en-dens}), (\ref{lambda-B}), and (\ref{helicality}), respectively.

In what follows we will be interested in the axial-dominated coupling $\kappa_{2}>\kappa_{1}$ since it allows to generate helical magnetic fields which have a better chance to survive in the post-inflationary evolution. For this case it is convenient to define a parameter that characterizes the gauge field production. Following Ref.~\cite{Anber:2010}, we introduce $\xi_{\mathrm{GF}}=\dot{I}_{2}/(2H I_{1})$. It is well known in the literature \cite{Anber:2006,Anber:2010,Barnaby:2012,Caprini:2014,Anber:2015,Ng:2015,Fujita:2015,Adshead:2015,Adshead:2016,Notari:2016,Domcke:2018eki,Cuissa:2018,Shtanov:2019,Shtanov:2019b,Sobol:2019,Kamarpour:2019,Sobol:2021,Domcke:2020zez} that significant gauge field production occurs only when this parameter is (much) greater than unity. There is, however, a constraint  due to the fact that the generated field should not induce large non-Gaussianities in the CMB power spectrum \cite{Guzzetti:2016}. This requires $|\xi_{\mathrm{GF}}|\lesssim 2.5$ at $N\sim 60$ $e$-foldings before the end of inflation, when the modes of perturbations relevant for CMB cross the Hubble horizon. A similar constraint ($|\xi_{\rm GF}|\lesssim 3$) is also derived in Ref.~\cite{Talebian:2021} from the requirement that backreaction is irrelevant. Although these are rather stringent constraints, we will not consider them in our analysis, assuming that there exists a mechanism to suppress gauge field production on CMB scales. E.g., this may occur due to Schwinger pair production which, however, lies beyond the scope of the present article. In the numerical analysis, we consider at most the last 40--50 $e$-foldings of inflation and  we focus on the properties of gauge fields generated at the end of inflation.

\subsection{The metric case}

In the metric case, the gauge-field production parameter $\xi_{\mathrm{GF}}$ has the form
\begin{eqnarray}
	\label{xi-full}
	\!\!\!\!\!\!\!\xi_{\mathrm{GF}}&=&\frac{\dot{I}_{2}}{2HI_{1}}\nonumber\\
	&\simeq& -\frac{2n}{3}\!\cdot\!\frac{2\kappa_{2}\left(6c_{0}\right)^{n}e^{\sqrt{\frac{2}{3}}\frac{\phi}{M_{p}}}\left[e^{\sqrt{\frac{2}{3}}\frac{\phi}{M_{p}}}-1\right]^{n-2}}{1+2\kappa_{1}\left(6c_{0}\right)^{n}\left[e^{\sqrt{\frac{2}{3}}\frac{\phi}{M_{p}}}-1\right]^{n}},
\end{eqnarray}
where $c_{0}\equiv \frac{12\pi^{2} A_{s}}{N_{\ast}^{2}}\approx 10^{-10}$  and the ``$\simeq$'' sign means that we used the slow-roll approximation in order to derive this expression.

For convenience  we   express the parameter $\xi_{\mathrm{GF}}$ in terms of number $e$-foldings $N$,
taking into account that for potential (\ref{pot-eff-single})
inflation ends at $\phi_{e}\approx 0.94 M_{p}$
and the field is trans-Planckian during the whole inflation stage, so that $e^{\sqrt{\frac{2}{3}}\frac{\phi}{M_{p}}}\approx \frac{4}{3}N$  for 
$\phi \gg \phi_{e}$: 
\begin{equation}
	\label{xi-full-N}
	\xi_{\mathrm{GF}}\simeq -\frac{2n}{3}\cdot\frac{2\kappa_{2}\left(6c_{0}\right)^{n}\frac{4}{3}N
	\left[\frac{4}{3}N-1\right]^{n-2}}{1+2\kappa_{1}\left(6c_{0}\right)^{n}\left[\frac{4}{3}N-1\right]^{n}},
\end{equation}

For $\kappa_{1}\left[e^{\sqrt{\frac{2}{3}}\frac{\phi}{M_{p}}}-1\right]^{n}\ll \left(6c_{0}\right)^{-n}$ (or equivalently $\kappa_{1}N^{n}\ll \left(8c_{0}\right)^{-n}$), the first term in the denominator dominates and we have
\begin{eqnarray}
	\label{xi-asym-1}
	\xi_{\mathrm{GF}}&\simeq& -\frac{4n}{3}\kappa_{2}\left(6c_{0}\right)^{n}e^{\sqrt{\frac{2}{3}}\frac{\phi}{M_{p}}}\left[e^{\sqrt{\frac{2}{3}}\frac{\phi}{M_{p}}}-1\right]^{n-2}\nonumber\\
	&=&-\frac{4n}{3}\kappa_{2}\left(6c_{0}\right)^{n}\frac{4}{3}N
\left[\frac{4}{3}N-1\right]^{n-2},
\end{eqnarray}
which for case $N\gg 1$ is
\begin{equation}
	\label{xi-asym-2}
	\xi_{\mathrm{GF}}\simeq -n \kappa_{2} (8c_0)^n N^{n-1}.
\end{equation}

We will see below that for $n>1$ this dependence $\xi_{\mathrm{GF}}(N)$ allows to obtain a magnetic spectral index $n_{B}<4$, in some cases we find an almost scale-invariant and even a red-tilted spectrum. However, the decrease of the parameter $|\xi_{\mathrm{GF}}|$  with time also leads to a decrease in the energy density of the generated fields during inflation which is unfavorable from the point of view of magnetogenesis.

In the opposite case, when  $\kappa_{1}\left[e^{\sqrt{\frac{2}{3}}\frac{\phi}{M_{p}}}-1\right]^{n}\gg \left(6c_{0}\right)^{-n}$, the second term in the denominator of Eq.~(\ref{xi-full}) dominates and we obtain the following expression:
\begin{equation}
	\xi_{\mathrm{GF}}\simeq -\frac{\kappa_{2}}{\kappa_{1}}\cdot\frac{n}{6\sinh^{2}(\phi/\sqrt{6}M_{p})}=
	-\frac{\kappa_{2}}{\kappa_{1}}\cdot
	\frac{n}{2N}
	\end{equation}
This function  is always an increasing with time during inflation. Therefore it leads to the blue-tilted spectrum of the generated fields. 

\subsection{The Palatini case}

As we saw in Sec.~\ref{sec-HS}, in the Palatini formulation, it is convenient to introduce the dimensionless time $\tilde{t}=Ht$ in terms of which all equations become independent of the parameter $\xi_{s}$. Let us show that this rescaling is also possible in the gauge-field sector. Consider Eq.~(\ref{eq-mode-2}) and the boundary condition (\ref{Bunch-Davies-vacuum}). By introducing the rescaled momentum $\tilde{k}=k/H$, they can be rewritten in terms of the rescaled time $\tilde{t}$:
\begin{multline}
	\mathcal{B}^{\prime\prime}_{\lambda}(\tilde{t},\tilde{\mathbf{k}})+\mathcal{B}^{\prime}_{\lambda}(\tilde{t},\tilde{\mathbf{k}})\\
	+\!\bigg(\frac{\tilde{k}^{2}}{a^{2}}\!\!-\!\frac{I^{\prime}_{1}}{2I_{1}}\!-\!\frac{1}{\sqrt{I_{1}}}\frac{d^{2}\!\sqrt{I_{1}}}{d\tilde{t}^{2}}\!-\!\lambda\frac{\tilde{k}}{a}\frac{I^{\prime}_{2}}{I_{1}}\bigg)\mathcal{B}_{\lambda}(\tilde{t},\tilde{\mathbf{k}})=0\label{eq-mode-2-P}
\end{multline}
with the Bunch-Davies boundary condition
\begin{equation}
	\label{Bunch-Davies-rescaled}
	\mathcal{B}_{\lambda}(\tilde{t},\tilde{\mathbf{k}})=e^{-i\tilde{k}\tilde{\tau}(\tilde{t})}, 
	\qquad \tilde{k}\tilde{\tau}\to-\infty.
\end{equation}
Taking into account that the coupling functions $I_{j}=I_{j}(h)$ do not depend on $\xi_{s}$ and the time derivative
\begin{equation}
	I^{\prime}_{j}=\frac{\partial I_{j}}{\partial h} h'\approx -\frac{4M_{p}}{h}\frac{\partial I_{j}}{\partial h}
\end{equation}
also does not contain $\xi_{s}$, we confirm that the gauge-field sector can also be considered in terms of the rescaled quantities. In practice, this means that we can choose a fixed value of $\xi_{s}$, e.g., $\xi_{s}=0$, and compute the spectra of generated fields. For any other value of $\xi_{s}$ we can obtain the spectra by a simple rescaling of all momenta.

From Eqs.~(\ref{en-dens-sp-B}) and (\ref{en-dens-sp-E}) it is easy to realize that the energy densities of generated fields scale as $\rho_{E,B}\propto H^{4}$, the
helicity (\ref{helicity}) scales as $\mathcal{H}\propto H^{3}$, and the correlation length (\ref{lambda-B}) behaves as $\lambda_{B}\propto H^{-1}$. The inflaton energy density, in the same time, scales as $\rho_{\rm inf}\propto H^{2}$. The fact that different energy densities have different scaling laws has an important consequence: for a given set of coupling parameters $\kappa_{1,2}$, if  we have backreaction of the generated fields for $\xi_{s}=0$, we may find a value of $\xi_{s}$ leading to a smaller value of $H$ such that there will be no backreaction.

Again, let us consider the axial-dominated coupling $\kappa_{2}>\kappa_{1}$. The parameter $\xi_{\mathrm{GF}}$ which determines the intensity of magnetogenesis has the following form:
\begin{equation}
	\label{xi-full-P}
	\xi_{\mathrm{GF}}=\frac{I^\prime_{2}}{2I_{1}}\simeq -4n\cdot\frac{2\kappa_{2}c_{0}^{n}\Big(\frac{h}{M_{p}}\Big)^{2n-2}}{1+2\kappa_{1}c_{0}^{n}\Big(\frac{h}{M_{p}}\Big)^{2n}},
\end{equation}
or, taking into account that $\frac{h}{M_{p}}\approx \sqrt{8N}$ for case $h\gg h_{e}$ and $N\gg 1$,
\begin{equation}
	\label{xi-full-P2}
	\xi_{\mathrm{GF}}\simeq -n\cdot\frac{\kappa_{2}(8c_{0})^{n}N^{n-1}}{1+2\kappa_{1}(8c_{0})^{n}N^{n}},
\end{equation}

In the limiting case when the kinetic coupling is negligible,
\begin{equation}
	\xi_{\mathrm{GF}}\simeq -8n\kappa_{2}c_{0}^{n}\Big(\frac{h}{M_{p}}\Big)^{2n-2}=-n \kappa_{2} (8c_0)^n N^{n-1},
\end{equation}
while in the opposite case when the kinetic coupling is strong
\begin{equation}
	\xi_{\mathrm{GF}}\simeq -4n\frac{\kappa_{2}}{\kappa_{1}}\Big(\frac{M_{p}}{h}\Big)^{2}=-\frac{\kappa_{2}}{\kappa_{1}}\cdot \frac{n}{2N}.
\end{equation}
Note, that in the Palatini case, the dependence of the parameter $\xi_{\mathrm{GF}}$ on the number of $e$-foldings from the end of inflation is exactly the same as for the metric case (except maybe in a small region close to the end of inflation where the slow-roll analysis fails).

\subsection{Spectral index of the magnetic power spectrum}

One of the important characteristics of generated magnetic fields is the spectral index $n_{B}$ which determines the distribution of magnetic energy density among modes with different momenta. It is possible to estimate this quantity analytically in the case of purely axial coupling. In the two previous subsections it was shown that for both metric and Palatini formulations the dependence  of the parameter $\xi_{\mathrm{GF}}$ on the number of $e$-foldings $N$ far from the end of inflation are the same in both asymptotic cases $\kappa_{1}N^{n}\ll \left(8c_{0}\right)^{-n}$ or $\kappa_{1}N^{n}\gg \left(8c_{0}\right)^{-n}$,
and, when the kinetic coupling is negligible,  $\xi_{\mathrm{GF}}$ is given by
\begin{equation}
\label{xi-on-N}
    \xi_{\mathrm{GF}}=-C_{n} N^{n-1}, \qquad C_{n}=n\kappa_{2}(8c_{0})^{n}.
\end{equation}
This allows us to estimate quantitatively the enhancement of any mode which crosses the horizon sufficiently far from the end of inflation. Indeed, the mode crosses the horizon when
\begin{equation}
    \frac{k}{aH}=2|\xi_{\mathrm{GF}}|.
\end{equation}
Now, we use the fact that $H\approx \textrm{const}$ during Higgs-Starobinsky inflation and $a=a_{e}e^{-N}$, where bottom index ``$e$'' corresponds to the end of inflation. Taking into account Eq.~(\ref{xi-on-N}), we obtain the following relation:
\begin{equation}
    N_{k}-(n-1)\ln N_{k}=\ln\frac{2C_{n}a_{e}H}{k},
\end{equation}
where $N_{k}$ is the number of $e$-foldings  before the end of inflation when the mode with momentum $k$ crosses the horizon. For $N_{k}\gg 1$, this equation can be approximately solved as 
\begin{equation}
    N_{k}\approx \ln\frac{2C_{n}a_{e}H}{k},
\end{equation}
and 
\begin{equation}
\label{xi-on-N1}
    \xi_{\mathrm{GF},k}=\xi_{\mathrm{GF}}(N_{k})=-C_{n} \Big(\ln\frac{2C_{n}a_{e}H}{k}\Big)^{n-1}.
\end{equation}

As was  shown in Ref.~\cite{Durrer:2011} in the case $\xi_{\mathrm{GF}}\approx \mathrm{const}$, the spectrum of generated gauge fields can be found analytically. At the end of inflation it reads as
\begin{equation}
    P_{B}(k)\approx \frac{H^{4}}{4\pi^{2}}\Big(\frac{k}{a_{e}H}\Big)^{\!4}\cdot \frac{\sinh(2\pi|\xi_{\mathrm{GF}}|)}{2\pi|\xi_{\mathrm{GF}}|}.
\end{equation}
Although this expression was derived for the case of constant $\xi_{\mathrm{GF}}$, we may also apply it if this parameter is slowly varying during inflation. For this we have to replace $\xi_{\mathrm{GF}}$ with its value when the mode $k$ crosses the horizon $\xi_{\mathrm{GF},k}$. As it was shown in Ref.~\cite{Durrer:2011}, the enhancement of a given mode happens when it crosses the horizon. After that its amplitude is frozen and it does not change until the end of inflation. Then, for large $|\xi_{\mathrm{GF},k}|$, we then find  approximately the following  form of the magnetic power spectrum
\begin{equation}
    P_{B}(k)\approx \frac{H^{4}}{8\pi^{2}}\Big(\frac{k}{a_{e}H}\Big)^{\!4}\cdot \frac{\exp\Big[2\pi C_{n}\Big(\ln\frac{2C_{n}a_{e}H}{k}\Big)^{n-1}\Big]}{2\pi C_{n}\Big(\ln\frac{2C_{n}a_{e}H}{k}\Big)^{n-1}}.
\end{equation}
The magnetic spectral index can be computed as
\begin{eqnarray}
    n_{B}&\equiv&\frac{d \ln P_{B}(k)}{d \ln k}\nonumber\\
    &=&4\!-\!2(n-1)\pi C_{n}\Big(\ln\frac{2C_{n}a_{e}H}{k}\Big)^{n-2}\!\!\!+\!\frac{(n-1)}{\ln\frac{2C_{n}a_{e}H}{k}}\nonumber\\
    &\approx& 4\!-\!2(n-1)\pi C_{n}\Big(\ln\frac{2C_{n}a_{e}H}{k}\Big)^{n-2}, \label{nB}
\end{eqnarray}
where in the last step we assumed $n\geq 1$ and $\ln\frac{2C_{n}a_{e}H}{k}\gg 1$. Let us consider a few particular cases.

For $n=1$, we have $n_{B}=4$, i.e., the magnetic power spectrum is blue-tilted with the same slope as vacuum fluctuations. At any moment of time the energy density is dominated by the last modes which have just crossed the horizon. Consequently, the correlation length of the generated field is of the order of the horizon size at the end of inflation.

For $n=2$, Eq.~(\ref{nB}) gives $n_{B}=4-2\pi C_{2}=4-256\pi\kappa_{2}c_{0}^{2}=\mathrm{const}$. Thus, the spectrum is a power-law and the spectral tilt depends on the coupling parameter $\kappa_{2}$. By increasing it, one can reach a scale invariant ($n_{B}=0$ for $\kappa_{2}=1/(64\pi c_{0}^{2})$) or even a red-tilted spectrum ($n_{B}<0$ for $\kappa_{2}>1/(64\pi c_{0}^{2})$). The latter case is, however, more involved because in order to find the total generated field one has to introduce an infrared cutoff. For this, one would need to know the smallest value of $k$ which is enhanced or, in other words, the magnetic field will depend on the time when inflation started.

For $n>2$ the spectral index $n_{B}$ depends on $k$ -- it decreases with decreasing  momentum. For any given $\kappa_{2}$, close to the end of inflation $n_{B}$ approaches the vacuum value $n_{B}=4$, while it decreases for smaller momenta and finally becomes negative. Also in this case, an infrared cutoff is needed: Unless there is some mechanism cutting the growth of the spectrum in the infrared region, the magnitude of the generated field is divergent.

For $n=1$ and $n=2$ (for $\kappa_{2}<1/(64\pi c_{0}^{2})$) we can also estimate the magnitude of the generated magnetic energy density far from the end of inflation. For this we use the expression for $\rho_{B}$ derived in Ref.~\cite{Sobol:2019} for the case of large $|\xi_{\mathrm{GF}}|$:
\begin{equation}
    \rho_{B}=\frac{H^{4}}{224\pi^{3}}\cdot\frac{\exp(2\pi|\xi_{\mathrm{GF}}|)}{|\xi_{\mathrm{GF}}|^{5}}.
\end{equation}
Substituting here Eq.~(\ref{xi-on-N}), we obtain
\begin{equation}
    \rho_{B}(N)\approx\frac{H^{4}}{224\pi^{3}}\cdot\frac{\exp(2\pi C_{n}N^{n-1})}{C_{n}^{5}N^{5(n-1)}}.
\end{equation}
Below we present numerical results for the spectra of the generated fields which illustrate the qualitative results of this subsection.

\subsection{Numerical results for spectra and generated fields}

After this qualitative analysis let us also present numerical results. We considered two cases $n=2$ and $n=3$. For definiteness, we consider here only the Palatini formulation for the following reason. As  discussed above, the generation of gauge fields occurs in both formulations in a very similar way. However, in the Palatini case, we have an additional degree of freedom, the parameter $\xi_s$, which is not fixed by observations. This allows us to change the energy scale of inflation (which may help to avoid backreaction). First we analyze the spectra of the generated fields and their dependence on the coupling parameters $\kappa_{1,2}$.

\begin{figure}[ht!]
	\centering
	\includegraphics[width=0.95\linewidth]{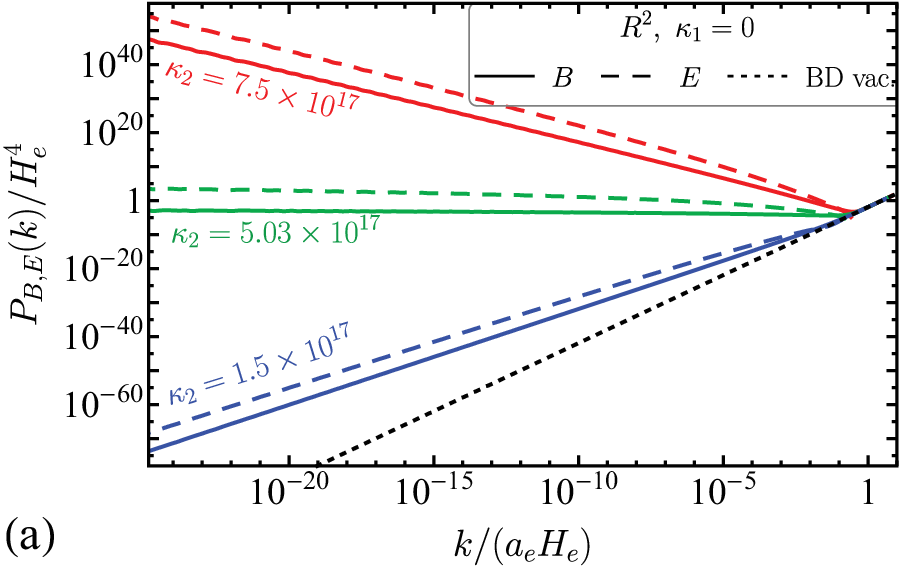}\\
\vspace*{2mm}
\hspace*{4mm}\includegraphics[width=0.89\linewidth]{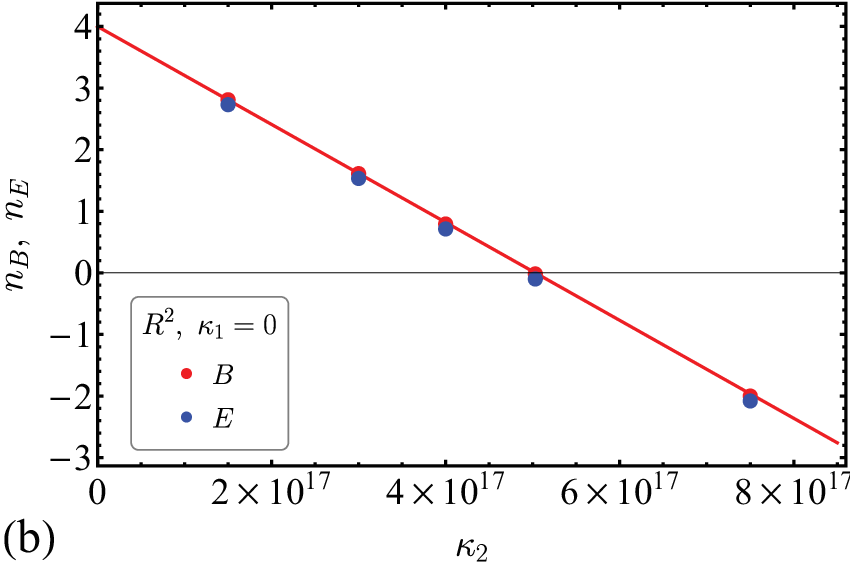}
	\caption{(a) Total magnetic (solid lines) and electric (dashed lines) power spectra at the end of Higgs-Starobinsky inflation in Palatini formulation with $R^{2}$-like nonminimal coupling of the gauge field. Blue, green, and red lines correspond to three values of the parameter $\kappa_{2}$: $\kappa_{2}=1.5\times 10^{17}$, $\kappa_{2}=5.03\times 10^{17}$, and $\kappa_{2}=7.5\times 10^{17}$, respectively. (b) Electric (blue dots) and magnetic (red dots) spectral indices as functions of $\kappa_2$ for $\kappa_1=0$. The red solid line shows the analytical estimate (\ref{nB}) for the magnetic spectral index.}
	\label{fig-sp-r2-ind}
\end{figure}

In Fig.~\ref{fig-sp-r2-ind}(a) we show the power spectra for $\kappa_1=0$ (purely axial coupling) and $n=2$. As discussed above, one can obtain blue, scale-invariant and red spectra, depending on the value of the coupling parameter $\kappa_2$. In Fig.~\ref{fig-sp-r2-ind}(b), the spectral indices $n_{B}$ and $n_{E}$ are shown as functions of $\kappa_2$. (The spectral indices are defined as $n_{B,E}=dP_{B,E}(k)/d\ln \, k$.) The analytical estimate (\ref{nB}) for the magnetic spectral index perfectly fits the corresponding numerical results. We emphasize that in the case of a red-tilted spectrum the total energy density is divergent in the infrared region and one needs to introduce some regularization in order to obtain finite results. Such a regularization can be easily obtained by introducing a nonzero kinetic coupling $\kappa_{1}\neq 0$.

\begin{figure*}[ht!]
	\centering
	\includegraphics[width=0.99\linewidth]{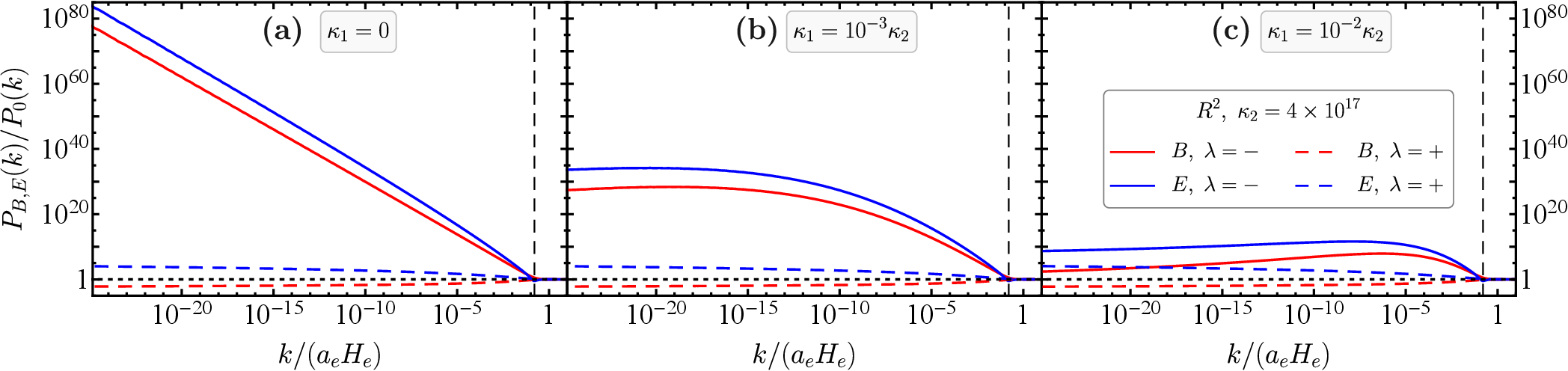}
	\caption{Magnetic (red lines) and electric (blue lines) power spectra for $\lambda=-$ (solid lines) and $\lambda=+$ (dashed lines) circular polarizations at the end of Higgs-Starobinsky inflation in Palatini formulation with $R^{2}$-like nonminimal coupling of the gauge field. The ratio of the power spectrum to the corresponding vacuum contribution $P_{0}(k)$ is shown [see Eq.~(\ref{P0})]. The value of the parameter $\kappa_{2}=4\times 10^{17}$ is fixed. Three panels correspond to three different values of the parameter $\kappa_1$: (a) $\kappa_1=0$, (b) $\kappa_{1}=10^{-3}\kappa_2$, and (c) $\kappa_1=10^{-2}\kappa_{2}$. The vertical dashed lines show the last mode which crosses the horizon during inflation.}
	\label{fig-sp-r2-grid}
\end{figure*}

In Fig.~\ref{fig-sp-r2-grid} we show the dependence of the power spectrum, in the case $n=2$ and fixed $\kappa_{2}=4\times 10^{17}$, on the value of $\kappa_{1}$. Panel (a) corresponds to the purely axial coupling and shows the spectrum of almost constant spectral index, as discussed above. Adding even a tiny kinetic coupling ($\kappa_{1}$ smaller than $\kappa_2$ by 2 or 3 orders of magnitude) drastically changes the shape of the spectrum in the infrared region, see panels (b) and (c). The spectral index becomes close to $n_B=4$, like for vacuum fluctuations. Note that the subdominant polarization does not change significantly when $\kappa_1$ increases (at least in the axial-dominant limit $\kappa_{1}\ll \kappa_{2}$) and its spectrum is always blue-tilted and close to the vacuum one (see the dashed lines in Fig.~\ref{fig-sp-r2-grid}).

\begin{figure*}[ht!]
	\centering
	\includegraphics[width=0.99\linewidth]{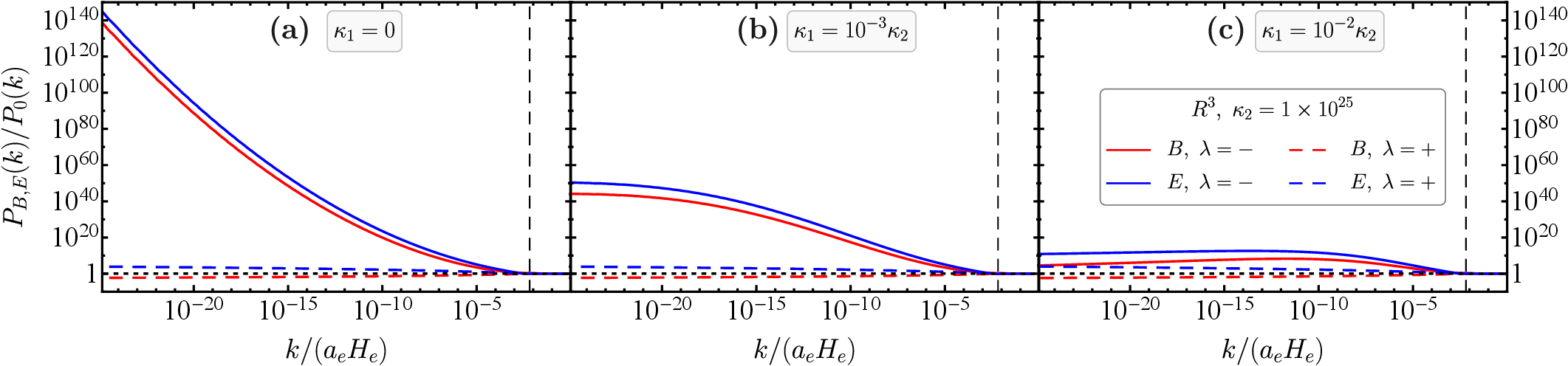}
	\caption{Magnetic (red lines) and electric (blue lines) power spectra for $\lambda=-$ (solid lines) and $\lambda=+$ (dashed lines) circular polarizations at the end of Higgs-Starobinsky inflation in Palatini formulation with $R^{3}$-like nonminimal coupling of the gauge field. The value of the parameter $\kappa_{2}=1\times 10^{25}$ is fixed. Three panels correspond to three different values of the parameter $\kappa_1$: (a) $\kappa_1=0$, (b) $\kappa_{1}=10^{-3}\kappa_2$, and (c) $\kappa_1=10^{-2}\kappa_{2}$. The vertical dashed lines show the last mode which crosses the horizon during inflation.}
	\label{fig-sp-r3-grid}
\end{figure*}

The power spectra for the case $n=3$ are shown in Fig.~\ref{fig-sp-r3-grid}. Again, panel (a) corresponds to the purely axial coupling. Here, as we discussed in the previous subsection, the spectral index is not constant, it decreases when we move to smaller momenta. Thus, for any $\kappa_2\neq 0$ we have the infrared problem and the total energy density (as well as correlation length, helicity etc.) is divergent. However, in full analogy with the case $n=2$, adding a small kinetic coupling resolves the problem. 


\begin{figure}[ht!]
	\centering
	\includegraphics[width=0.95\linewidth]{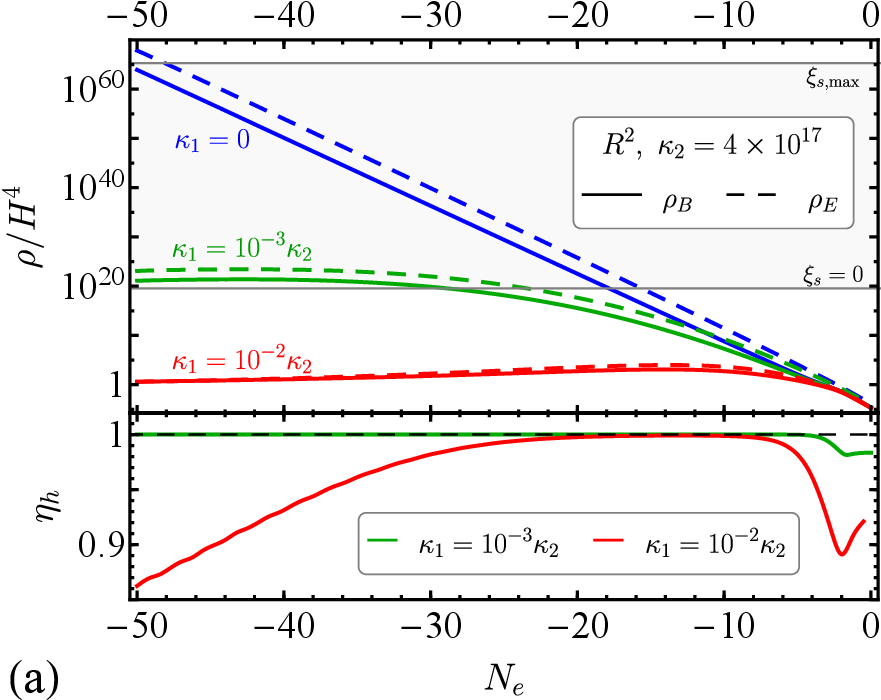}\\
	\includegraphics[width=0.95\linewidth]{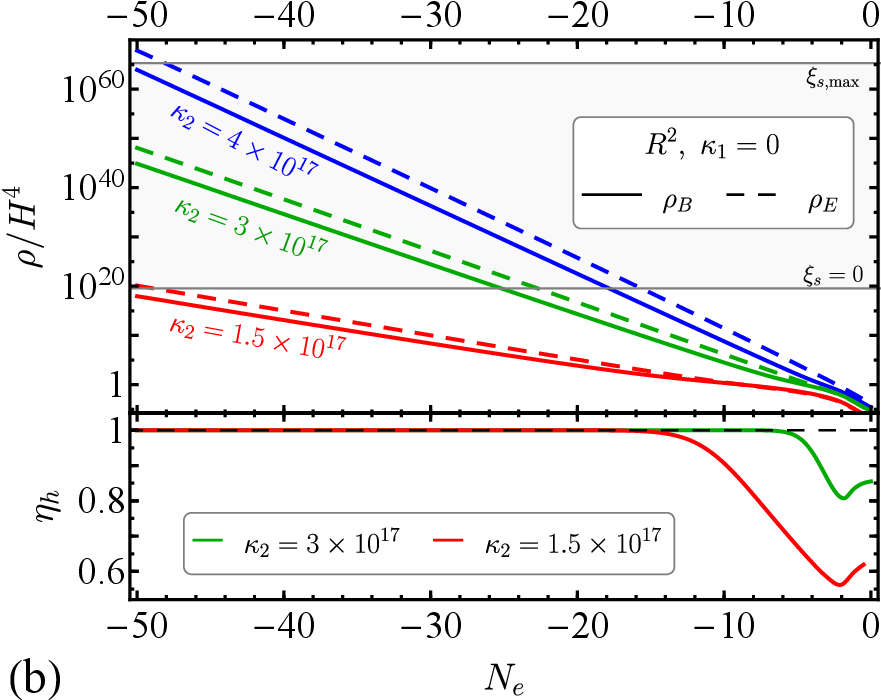}
	\caption{Top plots: magnetic (solid lines) and electric (dashed lines) energy densities as functions of the number of $e$-foldings before the end of Higgs-Starobinsky inflation in Palatini formulation with $R^{2}$-like nonminimal coupling of the gauge field. The values of the coupling parameters are the following: (a) $\kappa_{1}=0$ (blue lines), $\kappa_{1}=4\times 10^{14}$ (green lines), $\kappa_{1}=4\times 10^{15}$ (red lines), $\kappa_{2}= 4\times 10^{17}$ is the same for all curves; (b) $\kappa_{2}=4\times 10^{17}$ (blue lines), $\kappa_{2}=3\times 10^{17}$ (green lines), $\kappa_{2}=1.5\times 10^{17}$ (red lines), $\kappa_{1}= 0$ is the same for all curves. The gray shaded region on both panels shows all possible values of the inflaton energy density for different values of the parameter $\xi_{s}$ from 0 to the maximal value $\xi_{s,\mathrm{max}}$. Bottom panels show the corresponding values of helicality of the magnetic field, defined in Eq.~(\ref{helicality}).}
	\label{fig-rho-R2}
\end{figure}

\begin{figure}[ht!]
	\centering
	\includegraphics[width=0.95\linewidth]{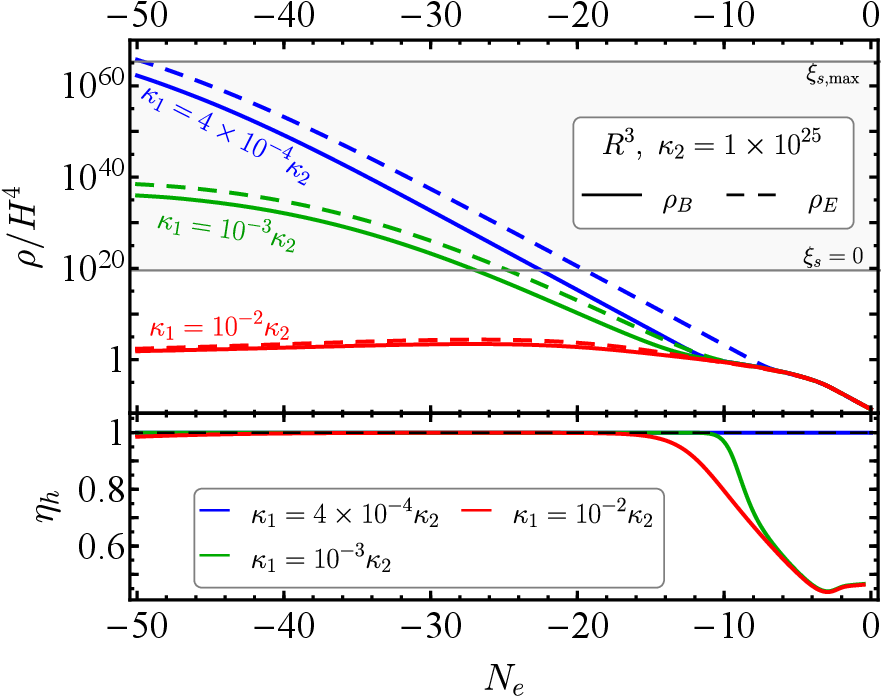}
	\caption{Top plot: magnetic (solid lines) and electric (dashed lines) energy densities as functions of the number of $e$-foldings before the end of Higgs-Starobinsky inflation in Palatini formulation with $R^{3}$-like nonminimal coupling of the gauge field. The values of the coupling parameters are the following: $\kappa_{1}=4\times 10^{21}$ (blue lines), $\kappa_{1}=10^{22}$ (green lines), $\kappa_{1}=10^{23}$ (red lines), $\kappa_{2}= 10^{25}$ is the same for all curves. The gray shaded region shows all possible values of the inflaton energy density for different values of the parameter $\xi_{s}$ from 0 to the maximal value $\xi_{s,\mathrm{max}}$. Bottom plot: the helicality of the magnetic field, defined in Eq.~(\ref{helicality}), for the same values of parameters.}
	\label{fig-rho-R3}
\end{figure}

Let us also consider the total energy density and helicity of the gauge field generated in the Higgs-Starobinsky model. The cases $n=2$ and $n=3$ are illustrated in Figs.~\ref{fig-rho-R2} and \ref{fig-rho-R3}, respectively. Plot (a), top panel in Fig.~\ref{fig-rho-R2} shows the magnetic (solid lines) and electric (dashed lines) components of the energy density for a fixed value of $\kappa_2=4\times 10^{17}$ and three values of $\kappa_{1}$. The value of $\kappa_{2}$ was chosen to be high enough to give strong generated fields but still without the infrared problem ($n_{B}\approx 0.8 >0$). A nonzero values of $\kappa_{1}$ strongly suppress the values of the generated fields at the beginning while close to the end of inflation the resulting values are pretty similar. Plot (b) , top panel in Fig.~\ref{fig-rho-R2} shows the same quantities for purely axial coupling at three different values of $\kappa_{2}$. Again all three values are such that $n_{B}>0$ and the infrared problem does not occur. The top panel in Fig.~\ref{fig-rho-R3} shows the energy densities of generated fields in the case $n=3$. Here, pure axial coupling always leads to a infrared problem, therefore, nonzero values of $\kappa_1$ were considered.

Typically, the energy densities are decreasing during inflation. Consequently, we should make sure that they do not cause  backreaction at least during the last 50-60 $e$-foldings of inflation when the perturbation modes, responsible for the CMB anisotropy observed today, cross the horizon. For this, we compare the energy density of gauge fields with that of the inflaton $\rho_{\rm inf}\approx 3H^{2}M_{p}^{2}$. In Fig.~\ref{fig-rho-R2}, we show the ratio $\rho/H^4$ and we compare it with the corresponding ratio for the inflaton, $\rho_{\rm inf}/H^{4}=3(M_{p}/H)^{2}$. In the Palatini case, by using Eq.~(\ref{Hubble-P}) this can be expressed as
\begin{equation}
\rho_{\rm inf}/H^{4}=36\Big(\frac{\xi_{h}^{2}}{\lambda}+\xi_{s}\Big).
\end{equation}
Obviously, the minimal value of this expression is achieved for $\xi_{s}=0$ (pure Higgs inflation). The maximal value is determined by the minimal possible value of the Hubble parameter during inflation. In this work, we choose it in such a way that the temperature after reheating (we assume it to be instantaneous) is not less than the electroweak scale, $T_{reh}\simeq 200\,\text{GeV}$. For definiteness, we use $\xi_{h}=10^{8}$, $\lambda=10^{-2}$. This implies $(\rho_{\rm inf}/H^{4})_{\rm min}\approx 3.6\times 10^{19}$ and $(\rho_{\rm inf}/H^{4})_{\rm max}\approx 2\times 10^{65}$. (The latter value corresponds to $\xi_{s,{\rm max}}\approx 5.4\times 10^{63}$.) These two values are shown by two gray lines in Figs.~\ref{fig-rho-R2}, \ref{fig-rho-R3}. The shaded region between these lines shows all possible values of the inflaton energy density which can be realized in our model. In order to avoid backreaction, the energy density of generated fields must be well below the chosen value of the inflaton energy density. The most efficient way to avoid the backreaction is to assume a nonzero kinetic coupling $\kappa_{1}$. However, as we have seen above, this leads to a blue-tilted spectrum and, consequently, to a smaller correlation length. Note that in all cases shown in the figures (without backreaction) the final value of the energy density is at least 20 orders of magnitude smaller than the inflaton energy density.

The bottom panels in Figs.~\ref{fig-rho-R2}, \ref{fig-rho-R3} show the helicality of the generated magnetic fields. [For the case $\kappa_1=0$ and $\kappa_2=4\times 10^{17}$ the result is not shown in Fig.~\ref{fig-rho-R2} because the spectral index $n_{B}=0.8<1$ and the corresponding integrals for helicity (\ref{helicity}) and the correlation length (\ref{lambda-B}) are divergent.] In the axially dominated case helicity is typically close to maximal, $\eta_{h}=1$. However, it is interesting to note that close to the end of inflation in some cases a deviation from unity is observed. This is because for those values of the coupling parameters $\kappa_{1,2}$ the spectrum is dominated by its UV edge, i.e., by the modes which cross the horizon close to the end of inflation. At this time, parameter $\xi_{\mathrm{GF}}$ is small and both circular polarizations have comparable magnitude. This implies that the generated fields are not maximally helical.

\begin{figure}[ht!]
	\centering
	\includegraphics[width=0.95\linewidth]{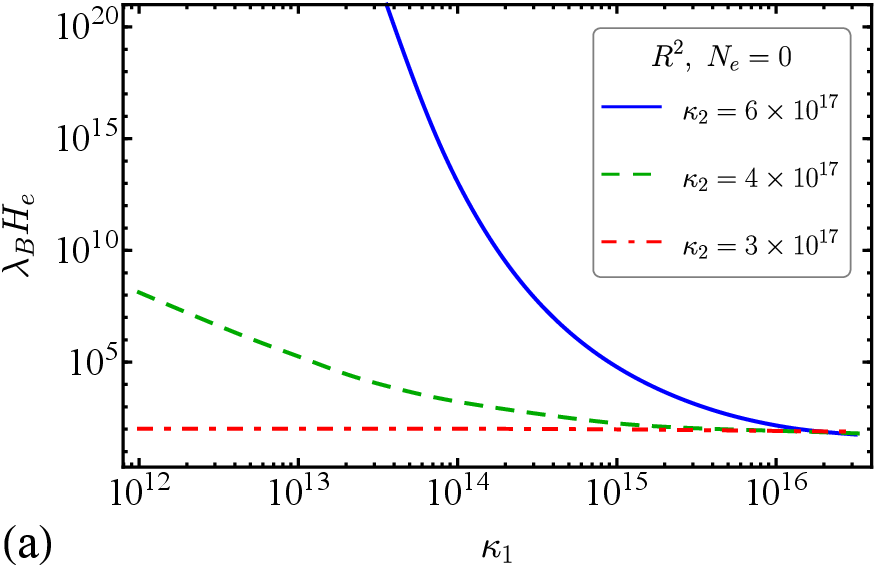}\\
	\includegraphics[width=0.95\linewidth]{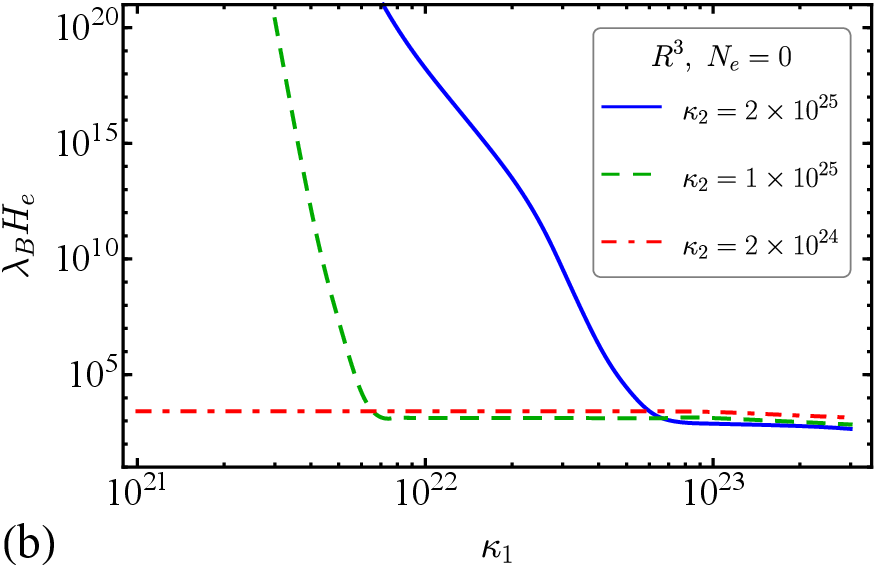}
	\caption{The correlation length of the generated magnetic fields at the end of Higgs-Starobinsky inflation in Palatini formulation as a function of the coupling parameter $\kappa_1$ for three fixed values of the parameter $\kappa_2$ which correspond to different lines. (a) Nonminimal coupling with $R^{2}$, $\kappa_{2}=6\times 10^{17}$ (blue solid line), $\kappa_{2}=4\times 10^{17}$ (green dashed line), $\kappa_{2}=3\times 10^{17}$ (red dashed-dotted line). (b) Nonminimal coupling with $R^{3}$, $\kappa_{2}=2\times 10^{25}$ (blue solid line), $\kappa_{2}=1\times 10^{25}$ (green dashed line), $\kappa_{2}=2\times 10^{24}$ (red dashed-dotted line).}
	\label{fig-lambda}
\end{figure}

Finally, let us discuss the correlation length of the generated fields. Its value at the end of inflation as a function of the coupling parameters $\kappa_{1,2}$ is shown in Fig.~\ref{fig-lambda}. There are three qualitatively different cases here: 
\begin{enumerate}
    \item[(i)] Both $P_B(k)$ and $P_{B}(k)/k$ are maximal close to the UV edge of the spectrum. Then, the correlation length is $\lambda_{B}\sim (10^{1}-10^{2}) H^{-1}$, i.e., is close to the horizon size during inflation.
    \item[(ii)] The power spectrum $P_{B}(k)$ is maximal at the UV edge of the spectrum, while the maximum of $P_{B}(k)/k$ is achieved for another, much larger mode. Then, $\lambda_{B}\sim (10^{4}-10^{8}) H^{-1}$.
    \item[(iii)] Both $P_B(k)$ and $P_{B}(k)/k$ are peaked far from the UV cutoff of the spectrum. The correlation length then can be arbitrarily large.
\end{enumerate}
Typically the case (i) is realized for small values of $\kappa_2$ (see the red dashed-dotted lines in Fig.~\ref{fig-lambda}) or for big $\kappa_1$. The case (iii) is always realized for $n_{B}<0$ and small $\kappa_{1}$ (see the blue solid lines in Fig.~\ref{fig-lambda}). The case (ii) is intermediate between the (i) and (iii). For $n=2$ it occurs for $0<n_{B}<1$ and small $\kappa_{1}$ [see the green dashed line in Fig.~\ref{fig-lambda}(a)].

\section{Conclusion}
\label{sec-concl}

The most popular models of inflationary magnetogenesis---kinetic and axial coupling models---suffer from a conceptual problem: there is an infinite number of ways to choose their coupling functions between the inflaton and gauge fields. Sometimes they are postulated without any physical motivation. In this paper, we present a model where the analytical form of the kinetic and axial coupling functions are deduced from a simple and physically well motivated setup. We consider the Higgs-Starobinsky inflationary model with an Abelian gauge field nonminimally coupled to a power of the scalar curvature given by the terms $\kappa_1 R^{n}F_{\mu\nu}F^{\mu\nu}$ and $\kappa_2 R^{n}F_{\mu\nu}\tilde{F}^{\mu\nu}$ in the Jordan frame. Switching to the Einstein frame, and treating these couplings perturbatively, we obtain nontrivial expressions for the coupling functions parametrized by just two real coupling parameters $\kappa_1$ and $\kappa_2$.

We considered two different formulations of gravity, the metric and Palatini ones. In the former formulation, it is well known that the Higgs-Starobinsky model reduces to an effective single-field inflation theory. Despite the fact that there are three parameters in this inflationary model (namely, the nonminimal coupling parameters $\xi_h$, $\xi_{s}$ and the Higgs quartic coupling constant $\lambda$), the inflationary dynamics is determined by a single combination of these parameters, $\xi_s+\xi_h^2/\lambda$. Moreover, this combination can be fixed by the normalization of the primordial scalar power spectrum from CMB observations. As a result, we obtain a model, which is completely equivalent to pure Starobinsky or pure Higgs inflation. On the contrary, in the Palatini formalism, the parameter $\xi_s$ remains free and determines the inflationary energy scale. However, by a proper rescaling of time variable, the Higgs-Starobinsky model in this case can be reduced to a purely Higgs model.

We have studied inflationary magnetogenesis in both metric and Palatini gravity formulations. It was  found that the dependence of the relevant parameter 
$\xi_{\mathrm{GF}}$ (which characterizes the magnetogenesis) on the number of $e$-foldings $N$ far from the end of inflation is the same in both formulations.

First, we considered magnetogenesis in the case of purely axial coupling ($\kappa_1=0$, $\kappa_2 \neq 0$). In this case, it is easy to estimate the behavior of the magnetic power spectrum. For $n=1$, it is blue-tilted with the corresponding spectral index $n_B=4$ (i.e., with the same slope as the spectrum of vacuum fluctuations). In this instance at any moment of time the energy density is dominated by the last modes which have crossed the horizon. Consequently, the correlation length of the generated field is of the order of the horizon size at the end of inflation. For $n=2$, the magnetic power spectrum also has a constant tilt, $n_{B}=4-2\pi C_{2}=4-256\pi\kappa_{2}c_{0}^{2}=\mathrm{const}$. 
By increasing the parameter $\kappa_2$, one may reach a scale invariant ($n_{B}=0$) 
or even a red-tilted spectrum ($n_{B}<0$). 
Finally, for $n>2$ the spectral index $n_{B}$ is not constant, it decreases with decreasing momentum $k$. For any given $\kappa_{2}$, close to the end of inflation, $n_{B}$ approaches the vacuum value $n_{B}=4$, while it decreases with  decreasing momentum and finally turns negative, red-tilted for very small momenta. In the limit $k\to 0$ the spectral density is divergent and, thus, the system has an infrared problem and one needs to provide some regularization in order to obtain physically meaningful results.

If we add a small kinetic coupling $\kappa_1\neq 0$, the spectrum becomes divided into two regions. For  Fourier modes which cross the horizon earlier than $N=1/(6c_0\kappa_1^{1/n})$ $e$-foldings before the end of inflation, the spectrum is always blue-tilted with $n_B\simeq 4$; however, for larger momenta the spectral index depends on $n$ and $\kappa_2$ in the same way as in the purely axial case discussed above. Therefore, a nonzero $\kappa_1$ performs the regularization and resolves the infrared problem.

Thus, we showed that it is possible to obtain a small spectral index $n_{B}$ (less blue-tilted spectrum) and, as a result, a larger correlation length of the generated fields. For this, one needs the gauge-field production parameter $|\xi_{\mathrm{GF}}|$ to be a decreasing function of time. Consequently, the generated fields will be decreasing as well. Provided that the backreaction problem is avoided during the whole inflationary stage, the final magnitude of the generated field appears to be very small.
Thus, in the $R^{n}$ nonminimal coupling model we have the following alternative: either we have a smaller spectral index $n_{B}<4$ and weak gauge fields at the end of inflation or sufficiently strong generated fields  but $n_{B}\simeq 4$.

If we want to achieve the best possible situation, i.e., keeping the large coherence length to increase the final magnitude of the generated field, we unavoidably face a backreaction problem. For instance, backreaction always occurs in the purely axial case $\kappa_1=0$ for $n>1$. In this situation, we must take into account the presence of gauge fields already at the moment when we perform the conformal transformation and switch to the Einstein frame. However, since we treat gauge fields perturbatively, we are not able to describe the backreaction regime. Nevertheless,  backreaction may strongly impact the inflationary dynamics. If it occurs around $N=50 - 60$ $e$-folding before the end of inflation, it may have a significant impact on the power spectra of primordial perturbations imprinted in the CMB anisotropy. Therefore, it would be interesting to study this regime in the full nonlinear picture. We plan to address this issue elsewhere.

\begin{acknowledgments}
	We would like to thank N. Shchutskyi and E. Gorbar for their participation at the initial stage of the work as well as  for useful discussions.
	The work of O.S. was supported by the ERC-AdG-2015 grant No. 694896 and by the National Research Foundation of Ukraine Project No. 2020.02/0062.
	The work of  R.D. and S.V. is supported by Swiss National Science Foundation Grant No.~SCOPE IZSEZ0 206908.
\end{acknowledgments}

\appendix

\section{THE LEGENDRE TRANSFORM  OF ACTION}
\label{app-particles}
In order to bring the action (\ref{a1}) to the canonical Einstein-Hilbert form, we have to  get rid of the quadratic term by performing a Legendre transform.
	With this aim we rewrite the action in the following form:
	\begin{equation}
		\!\!\!S[g_{\mu\nu},h]\!=\!\!\!\int\!\! d^4x \sqrt{-g} \Big[\!\!-\!\frac{M_p^2}{2}f(R,h)\!+\!\frac{1}{2}(\partial_\mu h)^{2}\!-\!\frac{\lambda}{4}h^4\Big],
	\end{equation}
	where
	\begin{equation}
		f(R,h)=\Big(1+\frac{\xi_h h^2}{M_p^2}\Big)R-\frac{\xi_s}{2M_{p}^{2}}R^2.
	\end{equation}
	We now introduce the new field
	\begin{equation}
		\Psi=\frac{\partial f}{\partial R}=1+\frac{\xi_h h^2}{M_p^2}-\frac{\xi_s R}{M_{p}^{2}}.
	\end{equation}
	From here,
	\begin{equation}
		R=\frac{M_{p}^{2}}{\xi_{s}}\Big(1-\Psi+\frac{\xi_h h^2}{M_p^2}\Big).
	\end{equation}
	Then, the Legendre transform of the function $f$ reads as
	\begin{equation}
		\label{F-function}
		F(\Psi,h)=\Psi R-f(R,h)=-\frac{M_{p}^{2}}{2\xi_{s}}\Big(1-\Psi+\frac{\xi_h h^2}{M_p^2}\Big)^{2}.
	\end{equation}
	Finally, formally writing the inverse Legendre transform $f(R,h)=\Psi R -F(\Psi,h)$, we obtain the action
	\begin{eqnarray}
	\label{action-Jordan-prep1}
	\!\!\!\!\!\!S[g_{\mu\nu},h,\Psi]&=&\int d^4x \sqrt{-g}\Big[-\frac{M_p^2}{2}\Psi R\nonumber\\
	&+&\frac{M_p^2}{2}F(\Psi,h)+\frac{1}{2}\partial_\mu h\partial^\mu h-\frac{\lambda}{4}h^4\Big],
	\end{eqnarray}
	which is now linear in $R$ but contains an additional scalar degree of freedom, $\Psi$, which is nondynamical, because the action does not contain a kinetic term for it.

 \end{document}